\documentclass[aip,showpacs,reprint,onecolumn,author-year,groupedaddress]{revtex4-1}

\usepackage{graphicx}
\usepackage{dcolumn}
\usepackage{bm}

\usepackage[english]{babel}    
\usepackage[T1]{fontenc}       
\usepackage[latin1]{inputenc} 
\usepackage{graphics,graphicx,color,psfrag,nicefrac,units}
\usepackage{amsmath,amsfonts,amssymb}
\usepackage[usenames,dvipsnames]{xcolor} \xdefinecolor{kurverot}{RGB}{153,61,113} \xdefinecolor{kurveblau}{RGB}{63,61,153} \xdefinecolor{kurvegold}{RGB}{153,140,61} \xdefinecolor{kurvegruen}{RGB}{0,178,0}
\newcommand{\bs}[1]{{\boldsymbol{#1}}}

\newcommand{\bk}{\bs{k}}\newcommand{\bq}{\bs{q}}

  \newcommand{\st}{\text{st}}       
 
\newcommand{\vct}[1]{\mathbf{#1}}
\newcommand{\epsU}{\varepsilon_\text{\tiny $U$}}

\begin{document}
\title{Transient stress evolution in repulsion and attraction dominated glasses}

\author{Christian P. Amann}\email{Christian.2.Amann@uni-konstanz.de}
\author{Matthias Fuchs}\email{Matthias.Fuchs@uni-konstanz.de}

\affiliation{Fachbereich Physik, Universit\"at Konstanz, 78457 Konstanz, Germany}

\date{\today}

\begin{abstract}
 We present results from microscopic mode coupling theory generalized to colloidal dispersions under shear in an integration-through-transients formalism. Stress-strain curves in start-up  shear, flow curves, and normal stresses are calculated with the equilibrium static structure factor as only input. Hard spheres close to their glass transition are considered, as are hard spheres with a short-ranged square-well attraction at their attraction dominated glass transition. The consequences of steric packing and physical bond formation on the linear elastic response, the stress release during yielding, and the steady  plastic flow are discussed and compared to experimental data from concentrated model dispersions.  
\end{abstract}

\maketitle

\section{Introduction}

 Colloidal dispersions have been established as model systems in materials science. They behave like fluids at high dilution, and form condensed phases if particle interactions dominate over entropic disorder. Equilibrium statistical mechanics explains equilibrium phases and their near-equilibrium properties based on direct particle interactions. While purely repulsive interactions lead to fluid and crystalline solids, attractions  can also give rise to liquids.  As colloidal dispersions offer the unique possibility to tailor the depth and range of the attractions, the conditions when liquid phases become stable could be established in studies  e.g.~by \cite{lekker92}.
 Yet colloidal dispersions  also form various {\em metastable solid states} like fractal networks, particle gels, and  glasses,  which can not be described with purely equilibrium statistical mechanics.  They are important in industrial processes and products, and in biological systems, which generally are far from thermal equilibrium. The specific mechanical properties of the resulting soft solids are often tailored adjusting the competition between attractive and repulsive interactions. Various solid states, with very different elastic properties, have been prepared; recent studies  include works by \cite{scio08}, \cite{schur11}, \cite{wagn13}, and others.

 Mixtures of colloids with non-adsorbing polymers  constitute one of the simplest systems, where effective interactions among the (larger) colloidal constituents can be adjusted  in a controlled way.  Two different glass states, viz.~amorphous solids formed by the freezing-in of the cooperative structural dynamics, could be observed conclusively by \cite{pham02} and \cite{bart02}, as had been predicted by \cite{fabb99} and  \cite{berg99} using mode coupling theory. Controlling the range of the attraction proved crucial for the formation of 'attraction driven glasses', where particles form physical bonds to their neighbors. Attraction driven glasses require short ranged attractions so that particles become tightly localized, as can be clearly seen in computer simulations \citep{puer02}. In 'repulsion driven glasses', the cooperative behavior of  the neighbors forming the cage localizes particles less tightly. 
 The localization length there corresponds to Lindemann's length \citep{lind10} which is roughly a tenth of the average particle separation \citep{hansen}. Lindemann had found that atomic displacements increase up to this value when (crystalline) solids are heated to their melting temperature. Mode coupling theory as developed by \cite{goetze} and others established  that Lindemann's length also characterizes the frozen-in structure of repulsion driven glasses, as was verified experimentally by \cite{puse87} and \cite{mege93} in colloidal hard sphere dispersions.   As attractions change the structure of these glasses only if they are of short range --- mode coupling theory predicts that qualitative changes require attraction ranges (somewhat) shorter than  Lindemann's length ---  and as attractions in molecular systems act across longer ranges, molecular glass transitions fall into the class of  repulsion driven transitions. 
 This has made hard sphere colloids a model system for studying the glass transition in molecular, metallic, and simple supercooled liquids, which was  recently reviewed by \cite{week12}. 
 Attraction driven glasses require attraction strengths of the order of only a few $k_BT$ (thermal energies) and can form at particle concentrations lower than repulsive ones  \citep{daws00}. The aspect that increasing the strength of attraction at first destabilizes and melts the (repulsion driven)  glass,  provides insight into the mechanism of vitrification and its theoretical description. For any strength, short ranged attractions increase the equilibrium probability of close-particle contacts, viz.~the contact value of the pair correlation function. Yet, this increased stickiness reduces the medium-ranged order in the disordered fluid as measured in the principal peak of the equilibrium structure factor. Mode coupling theory predicts that this decrease destabilizes the repulsion driven cages so that a hard sphere glass melts upon turning on a narrow attraction. 
 This causes an appreciable shift of the glass transition to higher concentrations. In a recent investigation, \cite{willenbacher} achieved a  shift of the transition packing fraction by more than 10 \%. For fixed concentration, short ranged attractions then cause a glass transition when at increasing attraction strength large wavevector contributions in the equilibrium structure factor contribute strongly. Inbetween, a reentrant fluid region lies where equilibrium is approached at long times.   The continuation of this states diagram to lower concentrations, where colloidal gels are observed \citep{scio08}, remains an active area of research. E.g.~computer simulations investigate it by following the temporal evolution of the average localization length of individual particles as function of concentration \citep{zacca09}. 

 Metastable glassy systems exhibit a complex interplay between their structural relaxation and flow causing a strongly non-Newtonian response. Moreover, the structural and mechanical changes during or after processing using flow or other external fields  also are crucial for achieving desired material properties. Thixotropy and ageing are important, and are moving into the reach of first principles theories only recently \citep{ball13}.  While a number of studies exists of the mechanical response of dispersions  approaching the repulsion driven glass transition --- a recent one \citep{ball09} where the transition density could be approached very closely is reviewed by \cite{sieb12} ---  there are far fewer studies of the nonlinear rheology in the complete region covering repulsion and attraction driven glass transitions. 
 Besides the already mentioned  work by  \cite{willenbacher} combining rheology and dynamic light scattering, especially the seminal investigations by \cite{pham06} and \cite{puse08} provided insights into the nonlinear mechanical behavior under different rheological protocols and varying repulsive and attractive effects by varying concentration and attraction strength. As also seen in computer simulations by \cite{puertas}, the linear shear moduli are far larger at the attraction driven glass transition than at the repulsion driven one. In mode coupling theory this arises from the dominance of large wavevectors in the approximated Green-Kubo relation for the linear moduli, which predicts an increase with the square of the inverse of the relative attraction range in the sticky limit \citep{berg00}. Another intriguing observation by Pham and coworkers concerns the yielding of glasses under applied strain. 
 A typical strain of around 10\% to 20 \% characterizes the shear-induced yielding of repulsion driven glasses when experiencing step strain and start-up flow, while attraction driven glasses yield in a two-step process. The former observation, in agreement with light scattering studies under large amplitude oscillatory shearing by \cite{puse02}, nicely ties to the picture of the cage effect and its characteristic length following Lindemann's criterion; yet this connection has not been established theoretically up to now. The second observation, already visible in oscillatory shear experiments by \cite{gada80} and supported by a detailed investigation under start-up shear flow by \cite{koum11}, can not be interpreted so easily by the cage picture because the two characteristic strain values are around 10\% and 100 \%. The second value is far larger than the local cage picture would imply, and the attraction range, which was around 5\% in the experimental system, appears not to characterize the stress-strain 
relations.  

 Aim of the present contribution is to determine the stress-strain relations close to repulsion and attraction driven glass transitions from mode coupling theory (MCT) as generalized to sheared colloidal dispersions in the integration-through-transients (ITT) framework. We will consider the quintessential repulsive glass transition, viz.~the one in a hard sphere fluid, and a typical attraction driven one for a narrow square-well pair potential. Specifically, we will consider start-up shear flow with fixed shear rate and determine the transient shear stress as function of accumulated strain. In the generalization of the microscopic mode coupling theory developed by \cite{brad12} based on the work by \cite{fuch02},  the structural relaxation under arbitrary, homogeneous, and incompressible flows is deduced from the equilibrium structure factor so that caging and bond formation, as in the quiescent situation, can be discussed. 

 Our work bears similarity to the study by \cite{fuch09} where hard disks were considered in two dimensions. Here we present the first calculations within microscopic MCT-ITT for hard spheres in three dimensions and additionally consider attractive glasses in the second part. Our work also bears some similarity to the one by \cite{miya04}, who, however, concentrated on  fluid states under shear and on time-dependent fluctuations around the steady state. 
 We focus here on the transient dynamics of yielding glass states and their stationary, time-independent properties which are not accessible to the theory by \cite{miya04}. Our work also bears similarity to the study by \cite{priy14} in the present volume of the Journal of Rheology, who also consider the non-linear rheology of  repulsion and attraction dominated glass transitions. They use a simplified MCT-ITT, where shear deformations are isotropically averaged, which enables them to study wider variations in the attraction ranges and strengths than possible in our solution of MCT-ITT without additional approximations.

 In Sect.~\ref{sec::ITT} the pertinent equations of mode coupling theory are summarized. Section~\ref{overv} gives an overview of the studied systems and the glass states diagram, while   Sects.~\ref{sec::HS} and \ref{sec::gel} describe the results for hard spheres without and with square-well attraction, respectively.  Section \ref{concl} concludes with a comparison of the findings with experimental data.

\section{Nonlinear rheology with mode coupling theory\label{sec::ITT}}

 The integration-through-transients (ITT) formalism which generalizes mode coupling theory (MCT)  to driven systems provides a method to calculate the complete time evolution of a concentrated dispersion under homogeneous strain deformation. This yields more information than e.g.~just obtaining the steady state properties as the evolution from elastic to plastic response can be observed. The MCT-ITT approach by \cite{brad12} is presently restricted to incompressible flows and neglects hydrodynamic interactions. 
 We will consider start-up shear flows, which is a simple time-dependent deformation protocol where the former condition is obeyed, and high particle concentrations, where the dominance of structural correlations motivates our neglect of hydrodynamics. Solvent effects are presumed to  renormalize the hydrodynamic radii and short time diffusion coefficients. 


 Microscopic starting point is the Smoluchowski equation for interacting Brownian particles in a given shear flow. The particles' time evolution follows from affine motion with the flow and random motion causing non-affine displacements, both combined in the Smoluchowski operator:
 \begin{equation}
  \Omega = \label{eq:smol} \sum_{i=1}^{N}\frac{\partial}{\partial {\bf r}_i}\cdot\left[\frac{\partial}{\partial {\bf r}_i} - {\bf F}_i - \dot\gamma \, y_i\, \hat{\vct{x}} \right].
 \end{equation}
 The force on particle $i$ derives from a potential, where the chosen pair interaction will enter in later sections. Dimensionless units are used, where length, energy and time are measured in units of  particle diameter $d$, thermal energy $k_BT$, and $d^2/D_0$, respectively. The effect of shear relative to Brownian motion is measured by the bare P\'eclet number Pe$_0=\dot\gamma d^2/D_0$, which in these units agrees with the shear rate. Note that the Weissenberg number Wi$=\dot\gamma\tau$, with $\tau$ an intrinsic $\alpha$-relaxation time scale of a fluid, is also called (dressed) P\'eclet number Pe. Viscoelastic response can be observed at Pe$_0\ll1$ and $\text{Wi}\gtrsim1$, while the response for $\text{Wi}\ll1$ is the one of a Newtonian fluid.

 An equation of motion for a {\em transient} density correlator $\Phi_{\bf q}(t)$ encodes rapid local motion without structural decay, and elasticity and plasticity owing to structural rearrangements. The transient density correlator  $\Phi_{\bf q}(t)=\langle \rho_{\bf q}^* \rho_{{\bf q}(t)}(t) \rangle / N S_q$, is the correlation function built with density fluctuation, $\rho_{\bf q} = \sum_{j=1}^N\, \exp{\{ i {\bf q}\cdot {\bf r}_j\}} $. Their time evolution is given by the adjoint of the Smoluchowski operator from Eq.~\eqref{eq:smol}, $ \rho_{\bf q}(t) = e^{\Omega^\dagger t} \rho_{\bf q}$. In ITT, the average can be performed over the equilibrium Gibbs-Boltzmann ensemble, because the system is assumed to be in equilibrium initially. 
 Thus the normalization, giving $\Phi_{\bf q}(0)=1$, is done with the equilibrium structure factor $S_q$.  The {\em shear-advected} wavevector ${\bf q}(t)= (q_x,q_y-\dot\gamma t q_x,q_z)^T$ appearing in the definition accounts for the  affine particle motion with the flow, and gives $\Phi_{\bf q}(t)\equiv1$ in the absence of non-affine motion. Random motion, affected by the shear flow, causes $\Phi_{\bf q}(t)$ to decay. In MCT-ITT this is given by an equation of motion containing a retarded friction kernel which arises from the competition of particle caging and shear advection of fluctuations
 \begin{equation}\label{mct1}
  \dot{\Phi}_{\bf q}(t) + \Gamma_{\bf q}(t) \; \left\{ \Phi_{\bf q}(t)
  + \int_0^t dt'\; m_{\bf q}(t,t') \; \dot{\Phi}_{\bf q}(t') \right\}
  =0 \;.
 \end{equation}
 The initial decay rate contains Taylor dispersion as $\Gamma_{\bf q}(t) = q^2(t)/ S_{q(t)}$. The generalized friction kernel $m_{{\bf q}}(t,t')$  is an autocorrelation function of fluctuating stresses. Based on the insights of quiescent MCT as described by \cite{goetze}, it is approximated by a quadratic polynomial in the  density correlators
 \begin{eqnarray}\label{mct2} 
  m_{\bf q}(t,t') =  \int\!\!\! \frac{d^3k}{(2\pi)^3} \frac{n S_{q(t)}
  S_{k(t')}\, S_{p(t')}}{2 q^2(t)\; q^2(t')}\; V_{\bf q k p}(t)\,
  V_{\bf q k p}(t')  \; \Phi_{{\bf k}(t')}(t-t')\,  \Phi_{{\bf p}(t')}(t-t')
 \end{eqnarray}
 with abbreviation ${\bf p}={\bf q}-{\bf k}$, and $n$ the particle density. The vertex function is given by
 \begin{equation}\label{mct3}
  V_{\bf q k p}(t)=  {\bf q}(t)\cdot \left( {\bf k}(t) \, c_{k(t)} + {\bf p}(t)\, c_{p(t)} \right)
 \end{equation}
 where $c_k$ is the Ornstein-Zernicke direct correlation function, $c_k=(1-1/S_k)/n$. These equations of motion were derived in detail by \cite{cate09} using a Zwanzig-Mori projection-operator formalism together with mode coupling approximations. 
 The equilibrium structure factor, $S_k$, encodes the particle interactions and introduces the experimental control parameters like density and temperature. Again generalizing quiescent MCT to flow, the potential part of the stress $\sigma_{\alpha\beta}(\dot\gamma)=\langle \sigma_{\alpha\beta}\rangle^{(\dot\gamma)}/V$ is approximated assuming that stress relaxations can be computed from integrating the transient density correlations
 \begin{equation}
  \sigma_{xy}(t)=\dot\gamma\int_0^t dt' \int \frac{d^3k}{2(2\pi)^3}\left[\frac{k^2_xk_y(-t')k_y}{kk(-t')}\frac{ S'_k S'_{k(-t')}}{S^2_k}\right]\Phi^2_{\bk(-t')}(t'). \label{eq::stressstrainMic}
 \end{equation}
 Equation~\eqref{eq::stressstrainMic} is the central equation which we will evaluate for different systems. It  can also be used to derive a constitutive equation  for the shear stress of the form $ \sigma_{xy}(t) =  \dot\gamma\int^t_{0}dt'\,g_{xy}(t',\dot\gamma)$ with a {\it generalized shear modulus} 
 \begin{equation}
  g_{xy}(t,\dot\gamma) = \frac{1}{2}\int \frac{d^3k}{(2\pi)^3}\, \left[\frac{k_x^2 k_yk_y(t)}{k(t)k}\frac{S_{k(t)}' S_k'}{S^2_{k(t)}}\right]\, \Phi^2_{\vct{k}}(t). \label{eq:modmct}
 \end{equation}
 Here a shift of the integration variable was performed collecting the wavevector advection in a time-dependent vertex (the term in the square bracket) which is weighted with the square of the density correlator at wavevector $\bf k$. Without wavevector advection, $g_{xy}(t,\dot\gamma=0)$ recovers the quiescent MCT expression for the stress autocorrelation function \citep{goetze}; see a review by \cite{fuch10} for more discussions. For an ideal elastic solid, the modulus $g_{xy}$ would be constant and stress and accumulated strain $\dot\gamma t$ would be proportional. If $g_{xy}(t)$ does not depend on $\dot\gamma$ and decays on an intrinsic time scale $\tau$, the finite time integral over $g_{xy}(t)$ is the long time viscosity $\eta^0_{xy}$ of a Newtonian fluid, stress and shear rate are then proportional. This recovers Maxwell's model of linear response. Viscoelastic media exhibit a non-linear behavior in $\dot\gamma$, because of a $\dot\gamma$ functionality of $g_{xy}(t,[\dot\gamma])$. 
 MCT can obviously provide a microscopic description of such viscoelasticity. Equation~\eqref{eq:modmct} will be studied  in the following sections specifying different systems by implementing different equilibrium structure factors.

\section{Overview of considered repulsive and attractive systems and their quiescent glass states}
\label{overv}

 In MCT-ITT, the   structural relaxation is determined by the direct correlation function $c_q$ (equivalently the  equilibrium structure factor $S_q$) which describes the effective interaction between density fluctuations. In density functional theory it gives the quadratic term in the interaction part of the free energy functional, in MCT it enters the interaction vertices when fluctuating stresses are connected to density fluctuations.   As discussed in detail by \cite{daws00}, there are two major mechanisms for vitrification in the self-consistency equations of MCT for pair potentials consisting of an excluded volume core and a short-ranged attraction.  They can be discerned from the wavevector range where the major contributions arise in the memory kernel of Eq.~\eqref{mct2}.

 \begin{figure}[htb]
  \centering
  \includegraphics[width=.7\linewidth]{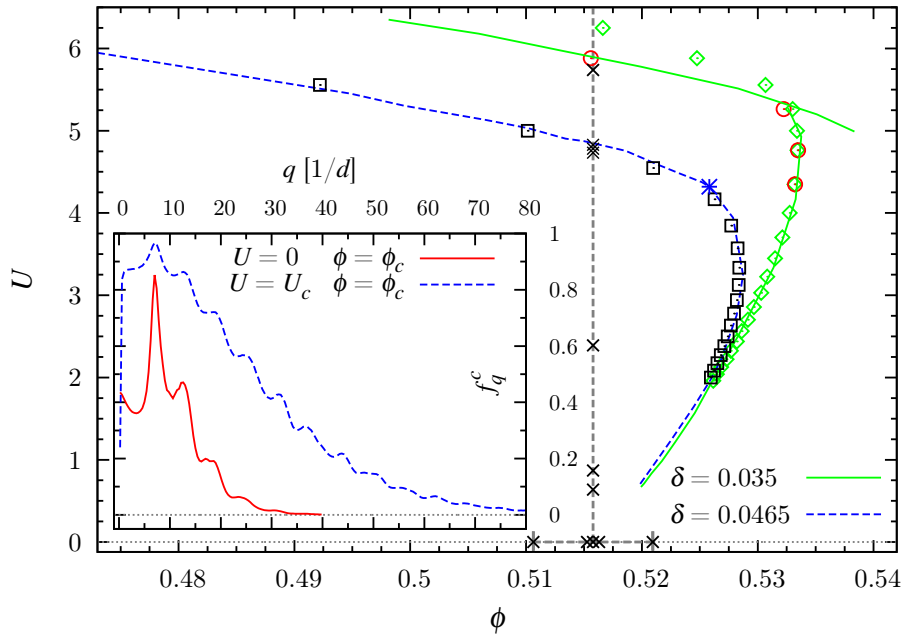}
  \caption{{\it Main panel:} States diagram showing the critical glass-transition interaction strength $U_c=u_0/k_BT$ and packing fraction $\phi_c$ for hard spheres with square-well attraction. The {\it solid lines} are taken from \cite{daws00}; see the legend for the relative attraction range of the SWP $\delta=\Delta/d$, with $\Delta$ attraction range and $d$ particle diameter. A {\it red star} marks the higher order glass transition point $A_4$ at $U=U_{A_4}$. 
  The {\it symbols} mark ADG calculations, with $q_\text{max}=79.8$  ({\it black squares and green diamonds}) or $q_\text{max}=119.8$ ({\it red circles}).  Calculations (marked by symbols $\boldsymbol \times$) along two paths crossing a glass transition will be discussed latter: Crossing the HS transition increasing $\phi$ for  $U=0$, and crossing the ADG increasing $U$ at $\phi_c$. 
  The {\it inset} shows (isotropic) non-ergodicity parameters $f^c_q$ at the two transitions, viz.~at the critical packing fraction $\phi_c$ for $U=0$ (hard-sphere case) and $U=U_c$ (attraction driven transition).\label{figMSA::DawsonPhases}}
 \end{figure} 

 The normal situation is that the principal peak in $S_q$ dominates, and that the glass transition is crossed when it becomes high. This glass transition originates in the local ordering caused by the excluded volume constraints in dense fluids, the transition leads to a 'repulsion dominated glass' (RDG), and the width of the glass form factor $f_q$ in reciprocal spaces is set by the localization length following Lindemann's criterion. The inset of  Fig.~\ref{figMSA::DawsonPhases} shows the glass form factor  $f_q^c$ at the transition of a fluid of monodisperse  hard spheres (HS), which is the most simple example for this transition \citep{goetze}. 
 Solving Eqs.~\eqref{mct1} to \eqref{mct3} numerically at vanishing shear rate and using the Percus-Yevick approximation for the structure factor, the transition lies at packing fraction $\phi_c=0.515712(1)$, which is somewhat below the value $\phi_c=0.58$ established in experiment for slightly polydisperse hard sphere colloids (\cite{puse87}, \cite{week12}). 
 The present numerical result for $\phi_c$ and especially $f_q^c$ (e.g.~the small wiggles close to the second and third peaks) differ from more precise calculations reviewed by \cite{goetze} because the chosen discretization of the wavevector integrals in Eq.~\eqref{mct2} is optimized to handle the anisotropic distortions under shear, including with attractions. Presently we cannot chose a finer discretization, because a single stress vs strain curve (from Fig.\ref{fig::Stress3DFlow}) takes  90  hours running time on a modern CPU.    

 A second mechanism causes arrest of the structural relaxation when a short-ranged attraction is strong enough in a colloidal dispersion.
 A square-well potential (SWP) of width $\Delta$ and attractive depth $u_0$ is chosen here to exemplify this. It acts outside the excluded volume core. This study  extends the one by \cite{daws00} by including shear flow. The potential depth will be made dimensionless using thermal energy $k_BT$, i.e. $U=u_0/k_BT$. Relative attraction ranges $\delta=\Delta/d$ smaller than Lindemann's ratio, $\delta < 0.1$  are required which cause strong contributions in the memory kernel of Eq.~\eqref{mct2} at large wavevectors. 
 The 'attraction dominated  glass' (ADG) transition takes place when wavevectors of the order $k\sim 1/\Delta$ dominate. The resulting glass form factors $f_q$ extend to large wavevectors and their $q$-dependent width corresponds to a localization length of the order of the attraction range $\Delta$. 
 This is shown in the inset of Fig.~\ref{figMSA::DawsonPhases}, where the form factors $f_q^c$ are shown at the attraction driven glass transition at $U_c=4.7811(8)$ and $\delta=0.0465$ at the same packing fraction $\phi_c$ as the HS transition; the different widths at the HS and ADG transitions are apparent. 

 The complete glass transition lines (states diagram) in Fig.~\ref{figMSA::DawsonPhases} for two attraction ranges verify that the first effect when turning on a short-ranged attraction is a destabilization of the repulsion dominated glass. The glass transition line moves to higher packing fractions initially with increasing attraction strength $U$. 
 The effect is stronger for smaller  relative attraction range $\delta$. A reentrant fluid region emerges which extends up to a maximal packing fraction, around where the two glass transition lines merge or intersect. The scenario depends on the precise attraction range. For  $\delta=0.0465$ both transition lines merge in an $A_4$ singularity, while for shorter ranges both glass transition lines intersect at a crossing point, and the ADG ends at an $A_3$ singularity \citep{daws00}. 

 In the present study, the transient stress evolution close to two typical transitions of both kinds shall be explored. Numerical calculations will be performed for two paths, one varying the concentration at fixed (vanishing) attraction strength, and the other one at fixed packing fraction for increasing well depth. Equilibrium structure factors, which are the only input to the theory, will be taken from the work by \cite{daws00}. While the choice of $U=0$ is quite natural and leads to the pure HS system, qualitatively similar results would hold for all RDG transitions at fixed $U$ (much) smaller than the attraction strength of the $A_4$ singularity.  The second transition to the ADG is chosen for the same packing fraction as the HS transition in order to eliminate density dependent differences. 
 According to MCT qualitatively similar results would hold for all $\phi$ below the density of the  $A_4$ singularity. In order to achieve a clear time scale separation between the slow structural relaxation of interest and the fast local dynamics, small relative separations from the glass transitions shall be considered. They will be expressed using the relative separations $\varepsilon=(\phi-\phi_c)/\phi_c$ and $\epsU= (U-U_c)/U_c$, respectively.  

 \begin{figure}[htb]
 \includegraphics[width=\linewidth]{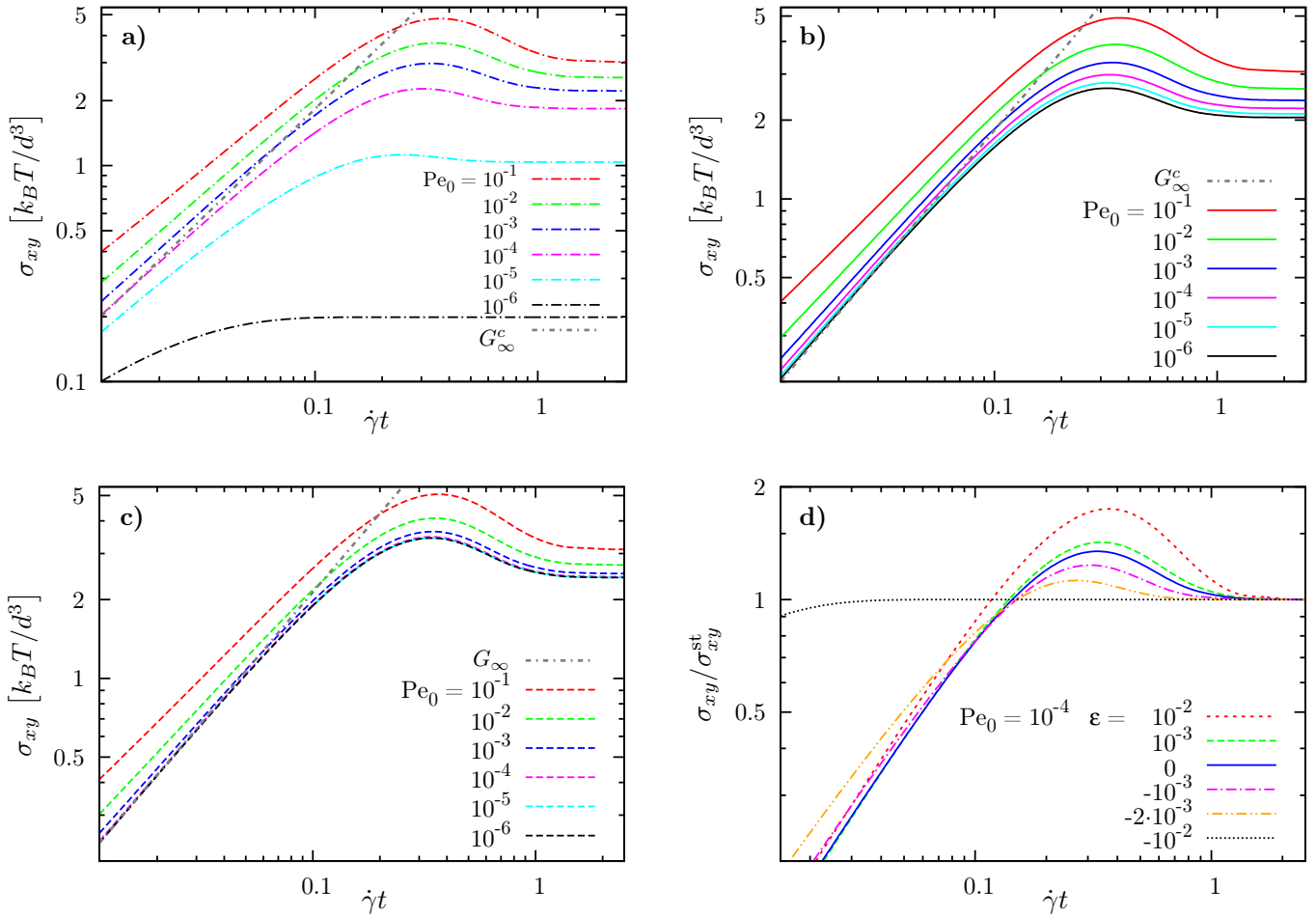}
 \caption[Stress vs strain curves in 3d MCT-ITT]{Transient shear stress $\sigma_{xy}$ as function of accumulated strain $\dot\gamma t$ for several separation parameters $\varepsilon=(\phi-\phi_c)/\phi_c$ and bare P\'eclet numbers Pe$_0$ calculated for a hard sphere system (see legends for~Pe$_0$). Panel {\it  a)} is in the fluid, $\varepsilon=-10^{-3}$,   {\it  b)} at the transition,  $\varepsilon=0^+$, and {\it  c)} in the glass,  $\varepsilon=10^{-3}$. The elastic moduli $G_\infty$ and $G_\infty^c$, calculated from the plateaus of Fig.~\ref{fig::modulg3D}, are shown as {\it gray, dash-dotted lines}. Panel {\it d)} shows stress curves rescaled by the steady state value varying  $\varepsilon$ at constant Pe$_0=10^{-4}$.\label{fig::strstnFG}}
 \end{figure}
 
\section{Hard-sphere transition\label{sec::HS}}

 After the presentation of the MCT equations and a short overview of the quiescent glass states diagram, the yielding of hard sphere glasses under applied shear strain shall be discussed first. It gives the example relevant to concentrated dispersions without especially short ranged attractions, and to molecular and metallic glass-forming liquids.

 Equation~\eqref{eq::stressstrainMic}  is a constitutive equation for the shear stress undergoing simple shear after start-up at $t = 0$. The vertex (the term in the square bracket) depends on time only via the accumulated strain, $\gamma=\dot\gamma t$. The squared correlators depend on time and accumulated strain independently in general and thus the stress-strain relations obtain different forms depending on the bare P\'eclet number Pe$_0$ and the distance to the glass transition $\varepsilon$. 
 The numerical calculation proceeds by first solving the self-consistency equations \eqref{mct1} to \eqref{mct3} for the density correlators $\Phi_{\bf q}(t)$ at given shear rate $\dot\gamma$, and second by integrating Eq.~\eqref{eq:modmct} to obtain the generalized shear modulus. If desired, flow-curves and stress strain relations can then be obtained. 

\begin{figure}[htb]
  \centering
  \includegraphics[width=.7\linewidth]{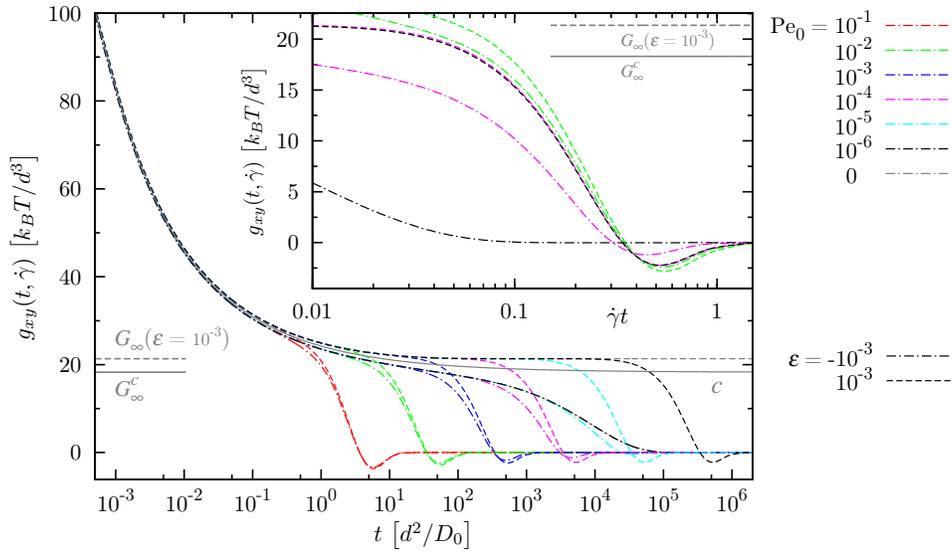}
 \caption[Generalized shear modulus in 3d MCT-ITT]{Generalized shear modulus $g_{xy}(t,\dot\gamma)$, Eq.~\eqref{eq:modmct}, as function of time {\it (main panel)} and accumulated strain ({\it inset}; sets for every second Pe$_0$ left out for clarity). The legend provides {\it color coded} the strain rates and {\it line-style coded} the separation parameters $\varepsilon$; a letter $\textcolor{gray}{c}$ labels the critical $g_{xy}(t,0)$ for $\varepsilon=0^+$. In the main panel, fluid curves at Pe$_0=0$ and $10^{-6}$ overlap, in the inset glass curves at Pe$_0=10^{-6}$ and $10^{-4}$ overlap. Elastic shear moduli $G_\infty(\varepsilon)$ can be read off from quiescent curves (Pe$_0=0$), with $G_\infty(0^+)=G_\infty^c=18.3\,k_BT/d^3$ and $G_\infty(10^{-3})=21.4\,k_BT/d^3$. 
 The {\it inset} shows subtle differences in the typical strain $\gamma_*$, where $g_{xy}=0$, when shear-driven and internal relaxation in $\Phi_{\bf q}(t,\dot\gamma)$ interfere. This becomes most clear for Pe$_0=10^{-6}$ in the fluid phase, where the internal relaxation dominates and the undershoot disappears.\label{fig::modulg3D}}
 \end{figure}

\begin{figure}[htb]
  \centering
  \includegraphics[width=.7\linewidth]{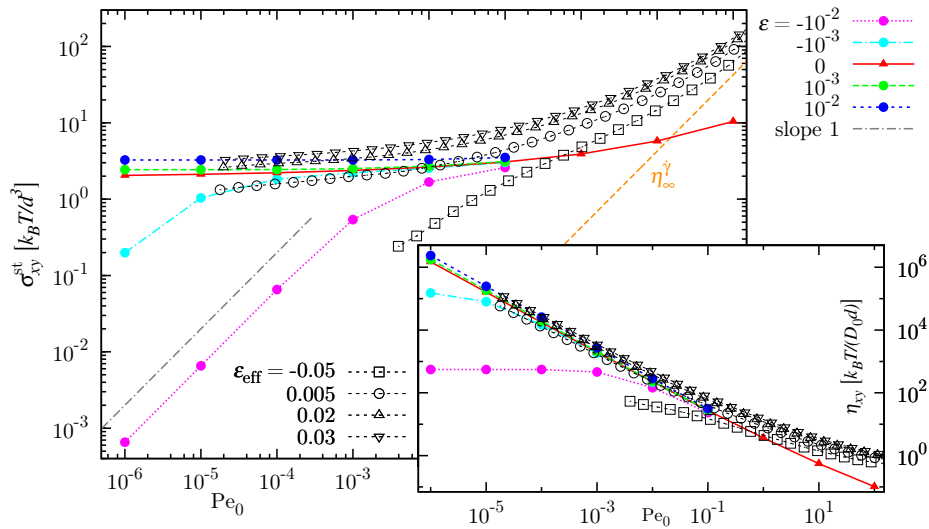}
  \caption[Flow curves and shear viscosities in 3d MCT-ITT]{The {\it main panel} shows flow curves $\sigma_{xy}(t\rightarrow\infty)=\sigma^\st_{xy}$ vs bare P\'eclet number Pe$_0$ for separation parameters $\varepsilon=(\phi-\phi_c)/\phi_c$ as given in the legend. The {\it colored  symbols} denote numerically obtained values for six shear rates Pe$_0=10^{\{-6;\ldots;-1\}}$ which are connected with straight lines as guides to the eye. For $\varepsilon=0^+$, higher Pe$_0$ were calculated up to the failure of the numerics at Pe$_0\ge10^{1}$; the Pe$_0=10^{2}$ point was computed without friction kernel in Eq.~\eqref{mct1} (i.e.\ just Taylor dispersion). 
  The open black symbols with effective separation parameters $\varepsilon_\text{eff}$ given in the legend are experimental data obtained by \cite{cras08}; the separation parameters $\varepsilon_\text{eff}$ were deduced from loss $G''(\omega)$ and elastic $G'(\omega)$ spectra. The measured effective packing fractions were $\phi_{\rm eff}=0.540, 0.580, 0.608$ and 0.622. A gray dash-dotted line illustrates a slope of 1 and thus the Newtonian regime.
  The high shear viscosity $\eta_\infty^{\dot\gamma}$ indicates the hydrodynamic contribution ($\eta_\infty^{\dot\gamma}= 1.56 k_BT/(D_0 d)$ as measured by \cite{cras08}) neglected in the MCT-ITT calculations. The comparison is discussed in  Sect.~\ref{concl}. The {\it inset} shows the stationary viscosity $\eta_{xy}=\sigma^\st_{xy}/\dot\gamma$.\label{fig::Stress3DFlow}}
 \end{figure}

 Figures \ref{fig::strstnFG} and \ref{fig::modulg3D} provide a good qualitative illustration of the transient stress regime, as well as already the quantitative data. Figure \ref{fig::strstnFG}  shows the shear stress vs strain. Figure \ref{fig::modulg3D}  shows the generalized shear modulus. Three regimes of viscoelasticity can be identified from the curves:\\
 $(i)$ First the stress grows linearly with strain proportional to an elastic shear modulus $G_\infty$. This expresses Hooke's law, $\sigma_{xy} = G_\infty \gamma$.  Figure \ref{fig::modulg3D}  shows that this corresponds to  a plateau in $g_{xy}(t, \dot\gamma )$ for intermediate times. 
 It  is reached after some shear-rate independent short-time relaxation, and holds up to the final decay time.  This plateau has the value  $G_\infty$. The index $\infty$ can be understood to refer to an infinite intrinsic relaxation time, which characterizes an elastic solid. In fluid states, the intrinsic final structural or $\alpha$-relaxation causes a decay of $g_{xy}(t,\dot\gamma=0)$, which softens the stress-strain relations. See Fig.~\ref{fig::strstnFG} a), where $\sigma_{xy}(t)$ becomes smaller at low shear rates, i.e. it becomes plastic because $\dot\gamma\tau<1$. There is a loss of memory in the system, causing deviations from the linear elastic  and leading to a viscous response. 
 At times long compared to the intrinsic $\alpha$-relaxation time $\tau$, a Newtonian viscosity is observed, $\sigma_{xy}(t\gg\tau)\to\eta^0_{xy}\dot\gamma$. Above the glass transition,  MCT predicts that the glass structure  is persistent and that the plateau does not decay (ideal glass), which is an idealization which is not observed experimentally. 
 Nevertheless, the time-window where the shear modulus is nearly time-independent can be made arbitrarily large by supercooling further. In MCT, the frozen-in glass structure is the reason for solid elasticity. Increasing the shear rate, the shear-distorted structure of the hard spheres inside their yielding structural cages stores additional stress. This increase of $\sigma_{xy}(\gamma)$  at small strains is an 'anelastic' effect in the  $\beta$-process of MCT, which describes the instability of the cages trapping the particles \citep{voigt12}; this can be seen best in panels b) and c) of Fig.~\ref{fig::strstnFG}.\\
 $(ii)$ Shear induced decay melts the glass, viz.~the $G_\infty$-plateau decays after a decay time proportional to the inverse of the shear rate; see Fig.~\ref{fig::modulg3D}. Consequently the time integral in Eq.~\eqref{eq::stressstrainMic}, viz. the area under $g_{xy}(t)$ vs time, does not increase anymore. A steady state value of $\sigma_{xy}(t)$ is reached in Fig.~\ref{fig::strstnFG} at strains of order one,
 denoted as flow curve value $\sigma^{st}_{xy}(\dot\gamma)$; see Fig.~\ref{fig::Stress3DFlow} for the flow curve. 
 The approach to a steady flow curve requires plastic effects, which cause the final decay of $g_{xy}(t)$. For increasing Pe$_0$, the plateau decay starts earlier and the short-time relaxation becomes more important.\\
 $(iii)$ Between the steady-state regime of the flow curve and the elastic regime occurs a transient plastic regime. From Fig.~\ref{fig::modulg3D} can be verified that $g_{xy}(t)$ takes negative values prior to its final decay to zero. This negative area under the $g_{xy}(t)$ curve adds a negative portion to $\sigma_{xy}(t)$ so that the stress decreases onto the steady-state plateau. 
 The emerging bump in Fig.~\ref{fig::strstnFG} is called stress overshoot. The stress is maximal for the peak strain value $\gamma_*$, i.e. Eq.~\eqref{eq::stressstrainMic} identifies $\gamma_*$ as the zero of $g_{xy}(\gamma=\dot\gamma t)$. If, in the fluid phase, the structural relaxation time $\tau$ is smaller than the shear induced one, the stress overshoot vanishes; Fig.~\ref{fig::strstnFG}a) illustrates this. This agrees with a result from linear response theory, viz. that the equilibrium shear modulus $g_{xy}(t, \varepsilon < 0)$ is completely monotone \citep{goetze}, and is observed in experiments on colloidal dispersions by \cite{pete12}, \cite{koum12}, and \cite{aman13}. 

 Figure \ref{fig::Stress3DFlow} shows the flow curve values $\sigma^{st}_{xy}(\dot\gamma)=\sigma_{xy}(t\to\infty,\dot\gamma)$ of the steady-state regime of the stress-strain curves from Fig.~\ref{fig::strstnFG} and also the long-time shear viscosity $\eta_{xy} = \sigma^{st}_{xy}/ \dot\gamma$. The difference between fluid phase and glass phase in the context of MCT can be seen. 
 In the glass phase and for Pe$_0 \to 0$, the area under $g_{xy}(t)$ becomes proportional to $1/\dot\gamma$. In consequence, a constant dynamic yield stress $\sigma^+_{xy}$ can be read off directly from the flow curve for vanishing shear rate, $\sigma^+_{xy}=\sigma^{st}_{xy}(\dot\gamma\to0,\varepsilon\ge0)$. This stress is necessary to keep the glass yielding and flowing at infinitesimally small shear rates. Because this yield stress is non-zero, the Newtonian viscosity diverges in the glass. Increasing the shear rate, processes on intermediate time scales (including the $\beta$-process in MCT studied by \cite{goetze}) cause the stress to become larger. 
 The $\beta$-process is slowest close to the glass transition, and therefore the flow curve at $\varepsilon=0$ varies sensitively with shear rate already at very small  Pe$_0$. For a schematic model of MCT-ITT, \cite{hajn09} deduced  a Herschel-Bulkley law at the transition, which unfortunately cannot be tested in our calculations because of the coarse discretization. Deeper in the glass, the flow curve stays rather constant, because the $\beta$-process has become faster, and $\sigma^+_{xy}$ can be observed for a wider window in  Pe$_0$. 
 It must be distinguished from a static yield stress, which a static load would have to overcome to fluidize a glass. One could within this context try to identify the  peak-stress observed during the overshoot as static yield stress, but the corresponding shear-protocol is shear-rate and not stress controlled. 
 The static yield stress could be (and is) different if one increases the stress in a controlled manner until the glass starts to flow. The issue of creep does then arise, which can be modeled with a stress controlled MCT as shown by \cite{sieb12}.

 If in the fluid phase ($\varepsilon < 0$), the shear-rate independent structural decay characterized by the quiescent $\alpha$-time $\tau$ takes place much earlier than the shear-induced one at time $1/\dot\gamma$, the integral in Eq.~\eqref{eq::stressstrainMic} becomes shear-rate independent and leads to the Newtonian viscosity $\eta^0_{xy}$. The linear relation between stress and shear rate defines a Newtonian fluid, and more generally the Newtonian viscosity is defined in the limit Wi$=\dot\gamma\tau \to 0$. 
 A first Newtonian plateau can be identified for fluid states in Fig.~\ref{fig::Stress3DFlow} inset. A second Newtonian plateau would arise if the initial  decay $\Gamma_{\bf q}(t)$ in Eq.~\eqref{mct1} gave a rapid shear-rate independent decay of the transient correlators at high shearing. 
 This would also make $g_{xy}(t)$ independent from $\dot\gamma$  for high shear rates, viz. Pe$_0\not\ll 1$. The present MCT-ITT cannot address this, as it describes the physics of structural arrest at long times and uses for short-times the quiescent $S_q$ and Brownian motion with $D_0$ as input without further considering how shear might affect them prior to structural arrest. Another problem arises, because the actual numerical iteration algorithm becomes unreliable for Pe$_0>10$. The flow curve becomes non-monotonic for stronger shear rates, and can even turn negative for some parameter values. 

 How a stress overshoot emerges in microscopic MCT has been discussed already for an simplified model with isotropic shear-distortions by \cite{zaus08} and for schematic MCT by \cite{aman13}. It provides insights into the physical mechanisms involved in the yielding of glass. These will be discussed in more detail in context with Fig.~\ref{kintegral} comparing effects from repulsion and attraction. 
 Here, the pertinent results for hard spheres shall be summarized. The peak  in the transient stress indicates a characteristic strain value $\gamma_*$ for the yielding process of a glass because it only arises for $\text{Wi}\gg1$. Another quantity, the relative peak amplitude,  $\sigma^{pk}_{xy}/\sigma^{st}_{xy} - 1$, characterizes the stress built-up during the linear response of the glass, which is released during the later stage of the yielding process.
  
\begin{figure}[htb]
  \centering
  \includegraphics[width=.7\linewidth]{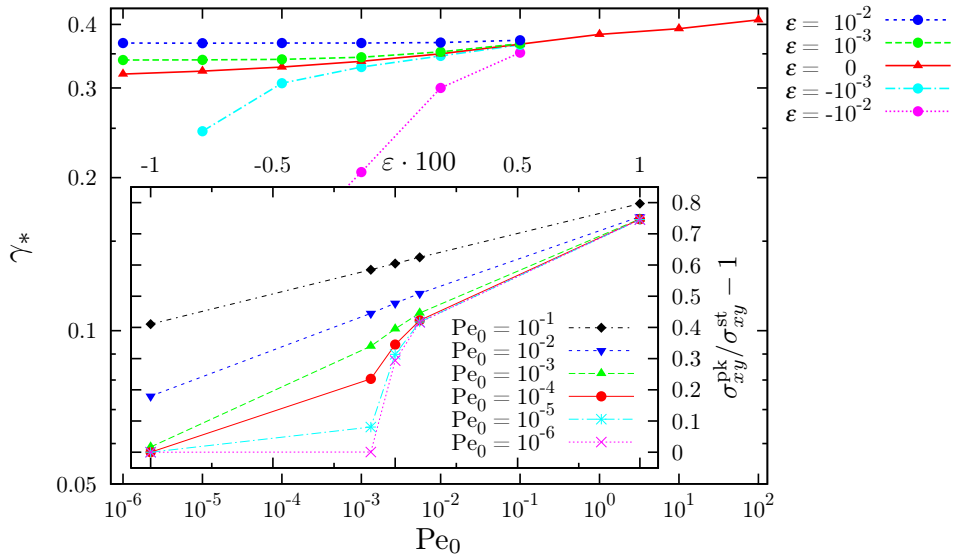}
  \caption[Peak strain in 3d MCT-ITT as function of Pe$_0$ and packing fraction]{The {\it main panel} shows peak-strain values $\gamma_*$ as function of bare P\'eclet number Pe$_0$. The grouping in separation parameters $\varepsilon$ is given by the outside legend.
  The $\gamma_*$s are read off from the zeros of $g_{xy}(t,\dot\gamma)$, Eq.~\eqref{eq:modmct}. {\it Symbols} are connected with straight lines as guide for the eye. The {\it inset} shows the relative overshoot height $\sigma^\text{pk}_{xy}/\sigma^\text{st}_{xy}-1$ of the MCT stress overshoots as function of separation parameters $\varepsilon$, grouped in bare P\'eclet numbers Pe$_0$; see inset legend. Both, peak-positions and peak-heights, increase with shear rate and with packing fraction.\label{fig::PeakAnal3D}}
 \end{figure}

 Numeric evaluations for the present hard sphere system, lead to the dependence of the peak strain $\gamma_*$, viz.~the zero of $g_{xy}(t,\dot\gamma)$, on shear rate and packing fraction, which is shown in Fig.~\ref{fig::PeakAnal3D}. 
 One verifies that in the glass ($\varepsilon > 0$) and for small Pe$_0$, the critical yield strain $\gamma_*$ is independent of shear rate. The reason is that intrinsic time scales then play no role in Eq.~\eqref{eq::stressstrainMic}, because the transient density correlators decay with accumulated strain, as does  the vertex in any case. 
 If the transient density correlators, which encode the structural relaxation, vary with time (and wavevector) even for negligible accumulated strain, then the value of $\gamma_*$ changes due to the dependence of the integral on the whole $\bf k$-range. When the linear response regime of the fluid dispersion is approached, the stress overshoot vanishes, so that the zero of $g_{xy}(t)$ still can be used to define  $\gamma_*$, but it loses its role as position of a noticeable overshoot in the stress-strain curve. 
 The inset in Fig.~\ref{fig::PeakAnal3D} shows the vanishing of the relative overshoot height for negative separation parameters,  $\varepsilon<0$.  Only if the the Weissenberg number Wi$=\dot\gamma\tau$ exceeds unity, a well developed stress overshoot allows to observe $\gamma_*$ in the stress-strain curve directly. For the fluid state at $\varepsilon= - 10^{-3}$, fitting a Kohlrausch law to the quiescent shear modulus $g_{xy}(t,0)$ gives $\tau=8063\,d^2/D_0$ (and stretching exponent $\beta=0.634$), so that Wi$=1$ holds at bare P\'eclet number Pe$_0=1.24\cdot10^{-4}$. This corresponds well to the inset in Fig.~\ref{fig::PeakAnal3D}, where one estimates a relative amplitude 0.238 (lin. interpolated) of the stress overshoot at this P\'eclet number. 
 Decreasing Pe$_0$ further, the overshoot vanishes quickly. Except for this rapid variation when the quiescent fluid is approached,  the position $\gamma_*$ of the stress overshoot increases smoothly with packing fraction for all shear rates. This can be seen from the main panel of Fig.~\ref{fig::PeakAnal3D}, and arises from the slight deepening of the undershoot in the  shear modulus $g_{xy}(t,\dot\gamma)$, which can be noticed in the inset of Fig.~\ref{fig::modulg3D}. The $\gamma_*$ also is an increasing function of shear rate for all separations to the glass transition.

 The overshoot's shape and height remains invariant, as long as internal relaxations play no role and parameters remain close to the glass transition, where $\tau$ is large. The inset of Fig.~\ref{fig::PeakAnal3D} shows that the relative overshoot height $\sigma^{pk}_{xy}/\sigma^{st}_{xy} - 1$, with $\sigma^{pk}_{xy}$ the maximal stress value, increases with packing fraction and with Pe$_0$ outside the asymptotic regime. The former effect is  due to the growth of the vertices and correlators in Eq.~\eqref{eq:modmct} with packing fraction, the latter due to the growing role of the $\beta$-decay with increasing shear rate. 
 Note that the increase with density differs from the findings by \cite{pete12}, where the opposite was measured. A decrease of the relative overshoot height with packing fraction was attributed to approaching random close packing (RCP). Besides speculations by \cite{aman13} if ageing played a role, this difference remains unresolved. It may indicate that  MCT-ITT does not describe the regime close to RCP, but around the glass transition, and that MCT's predictions fail at much higher or lower densities than at $\phi_c$.

\begin{figure}[htb]
  \includegraphics[width=\linewidth]{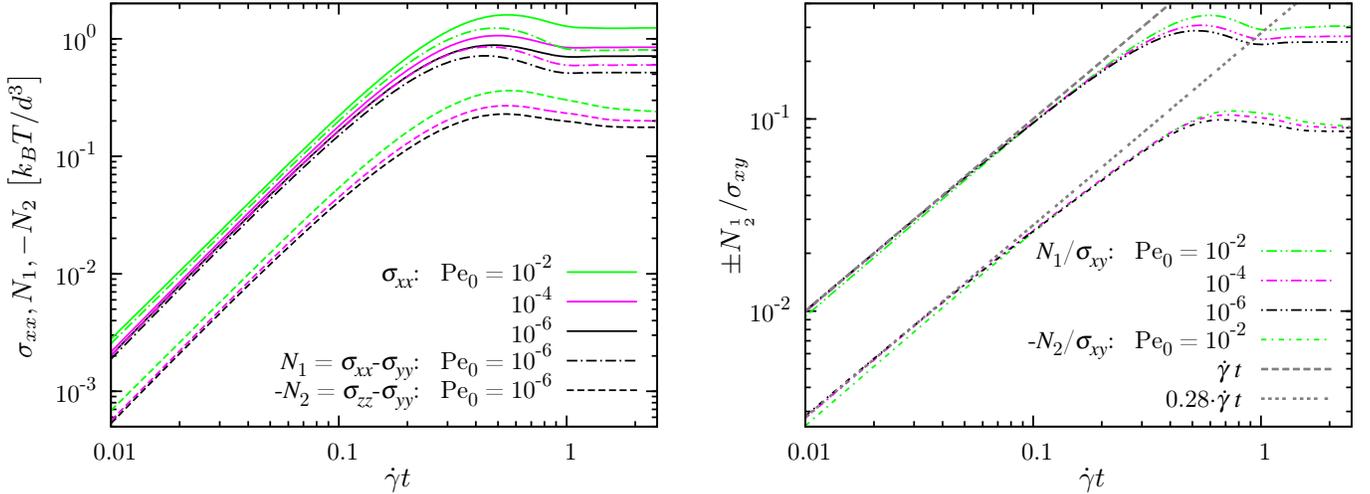}
  \caption[Normal stresses in 3d MCT-ITT]{{\it Left panel:} transient normal stresses $\sigma_{xx}$ ({\it solid lines}), {\it first normal-stress difference} ${N_1=\sigma_{xx}-\sigma_{yy}}$ ({\it dot-dashed lines}), and {\it second normal-stress difference} $N_2=\sigma_{yy}-\sigma_{zz}$ ({\it dashed lines}, plotted with negative sign) as functions of accumulated strain $\dot\gamma t$ at the critical packing fraction ($\varepsilon=0^+$) and for the Pe$_0$s given in the legend (Pe$_0$s are {\it color coded} for all {\it line-styles}). One verifies that all $\sigma_{ii}>0$ and $\sigma_{zz}>\sigma_{yy}$. Stress overshoots are apparent; they lie at 25\% higher peak strains than in the shear stress.\newline 
  The {\it right panel} shows the normal-stress differences divided by shear stress, $N_1/\sigma_{xy}$ and $-N_2/\sigma_{xy}$, to illustrate the proportionality of this ratio to the accumulated strain in the elastic region. {\it Grey dashed and gray dotted lines} are linear fits with prefactors 1 and 0.28 (see legend).\label{fig::Stress3DN1N2}}
 \end{figure}

 To illustrate the tensorial nature of the stress tensor $\boldsymbol{\sigma}(t)$ and of the 3d MCT-ITT approach, Fig.~\ref{fig::Stress3DN1N2} shows a calculation of the first and second normal-stress differences $N_1(t) = \sigma_{xx}(t) -\sigma_{yy}(t)$ and $N_2(t) =\sigma_{yy}(t) -\sigma_{zz}(t)$. Already \cite{brad09} showed tensorial, numerical evaluations of $\boldsymbol{\sigma}(t)$  via a schematic MCT. 
 However, in schematic MCT, $N_2 = 0$ under shear by construction.  More recently, \cite{brad13} calculated for small shear rates the leading quadratic order of both normal stresses for stationary flows in fluid states. Figure~\ref{fig::Stress3DN1N2} shows results for the nonlinear regime of yielding glasses complementing their study. 

 Within MCT-ITT under simple shear, all components of the stress tensor can be calculated analogously to Eq.~\ref{mct3}, by simply changing the components of some $\bf k$ vectors. Because of the neglect of an isotropic contribution \citep{cate09}, however, the shear-dependent pressure can not be calculated presently. Figure~\ref{fig::Stress3DN1N2} shows the stress on a plane-element perpendicular to the flow direction, viz.~$\sigma_{xx}(t)$, and the normal-stress differences as defined above, which are of distinguished rheological interest.
 The calculation has been done at the glass transition ($\varepsilon=0^+$) and for  Pe$_0 = 10^{-2;-4;-6}$, i.e. for a genuine, critical glass behavior with small Pe$_0$  and large Wi. The first observation is that they all exhibit a transient regime, with stress overshoots, which look qualitatively like those of the shear stress, cf. Fig.~\ref{fig::strstnFG}. The overshoots however occur at strains larger than 0.4, which is larger than all $\gamma_*$ determined from the shear stress.
 At all times $t > 0$, it holds $\sigma_{xx} > \sigma_{zz} > \sigma_{yy} > 0$, which renders $N_1 > 0$ and $N_2 < 0$.  Comparing more quantitatively, \cite{brad09} showed that the Lodge-Meissner relationship holds in the elastic regime. 
 It reads $N_1(t)/\sigma_{xy}(t) = \gamma$, and states that the slope of the 1st normal-stress difference is quadratic in accumulated strain in the elastic regime with prefactor $G_\infty$, the shear modulus. The full 3d numerics recovers this relation and gives a corresponding one for the 2nd normal-stress difference: $N_2(t)/\sigma_{xy}(t) \approx -0.28 \gamma$; both relations are shown in the right panel of Fig.~\ref{fig::Stress3DN1N2}. Outside the elastic regime, the simple linear relation between normal stresses and shear stress breaks down, as is clear from the different stress overshoot positions.

\section{Attraction driven glass\label{sec::gel}}

 Turning on a square-well potential (SWP) among the hard spheres, the attraction depth $U$ relative to the thermal energy becomes an additional control variable.  Increasing $U$ the states diagram depends on the relative attraction range $\delta$. We fix $\delta=0.0465$, far smaller than Lindemann's ratio, in order to explore the consequences of physical bond-formation in concentrated dispersions. At this $\delta$, the two different types of glass transitions, repulsion dominated glass (RDG) and attraction dominated glass (ADG), merge in an higher order singularity of MCT, denoted $A_4$ bifurcation point \citep{sperl02}. This occurs around $U_{A_4}\approx 4.3$; see Fig.~\ref{figMSA::DawsonPhases}. 
 For $U$ below this value, the major effect of the attraction is to destabilize the RDG and to induce a (reentrant) fluid phase where equilibrium is reached at long times (\cite{pham02} and \cite{bart02}). For attractions close to and slightly above  $U_{A_4}$ (and concentrations below it), the ADG transition takes place where the short-ranged attraction causes cooperative bonding among caging particles. Even though the attraction strength is of the order of a few $k_BT$ only, the 'physical bonding' stabilizes a second non-ergodic state with quite different mechanical properties than the RDG.   

 \begin{figure}[htb]
  \centering
  \includegraphics[width=.7\linewidth]{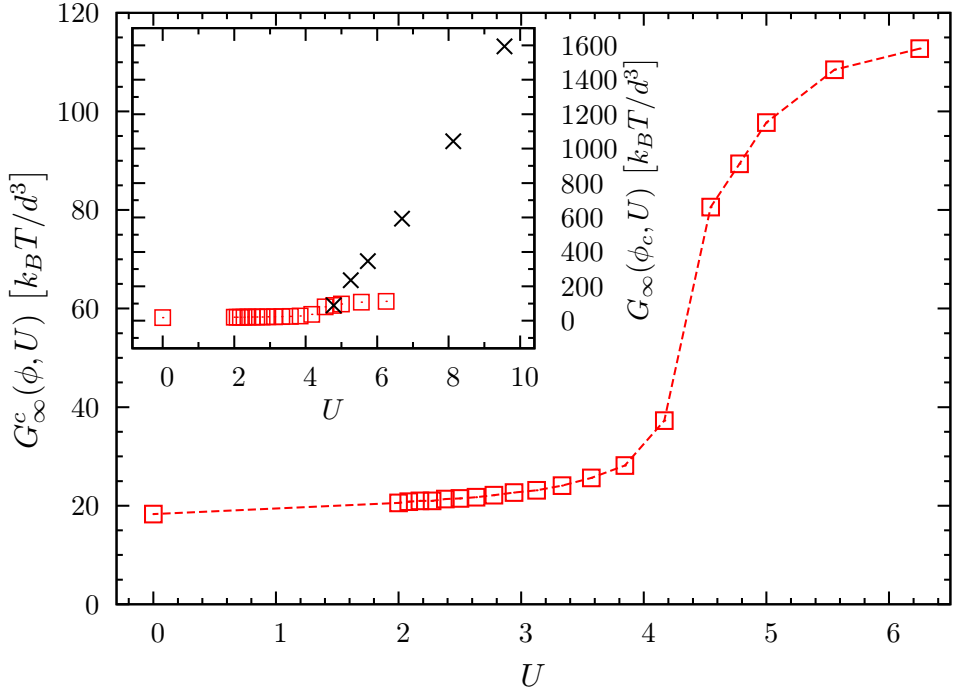}
  \caption{Shear moduli of hard spheres with square-well attraction of relative range $\delta=0.0465$ as function of attraction strength $U=u_0/k_BT$. Squares mark calculations along the glass transition line and correspond to the black squares in Fig.~\ref{figMSA::DawsonPhases}; the packing fraction $\phi_c$ varies non-monotonically here. The inset shows shear moduli varying over a wider range. Symbols $\boldsymbol\times$ mark $G_\infty$ calculated at fixed packing fraction $\phi_c$ and increasing $U$ deeper into the attraction driven glass. \label{moduln}}
 \end{figure}

 Figure \ref{moduln} shows the (quiescent) shear moduli $G_\infty^c$, viz.~the elastic coefficients under volume conserving deformations, along the glass transition lines as functions of attraction strength $U$; note that $\phi$ varies non-monotonically along the red curve in Fig.~\ref{figMSA::DawsonPhases} for these calculations. 
 For $U=0$ the modulus of HS indicated in Figs.~\ref{fig::strstnFG} and \ref{fig::modulg3D} is recovered. At first, the attraction little changes the elastic shear modulus relative to the RDG  value until around $U_{A_4}$, $G_\infty^c$ increases rapidly. For even shorter attraction range, e.g.~$\delta=0.03$, the modulus would jump discontinuously at the crossing of glass transition lines in Fig.~\ref{figMSA::DawsonPhases} \citep{daws00}. The mechanism causing the increase is the tighter localization of particles in the ADG than in the RDG. 
 It remains an entropic effect like in the RDG, as $G_\infty^c \propto k_BT$, but asymptotically for small ranges MCT predicts an increase scaling like: $G_\infty^{\rm c, ADG} \propto (1/\delta^2) G_\infty^{\rm c, RDG}$. Note that for increasing $U$ along the glass transition line (outside the range of Fig.~\ref{moduln}) the elastic modulus decreases non-monotonically, which was shown by \cite{berg03}. This happens because the particle concentration strongly decreases along the transition line (and with it the elastic modulus) for increasing interaction strength.

 \begin{figure}[htb]
 \includegraphics[width=\linewidth]{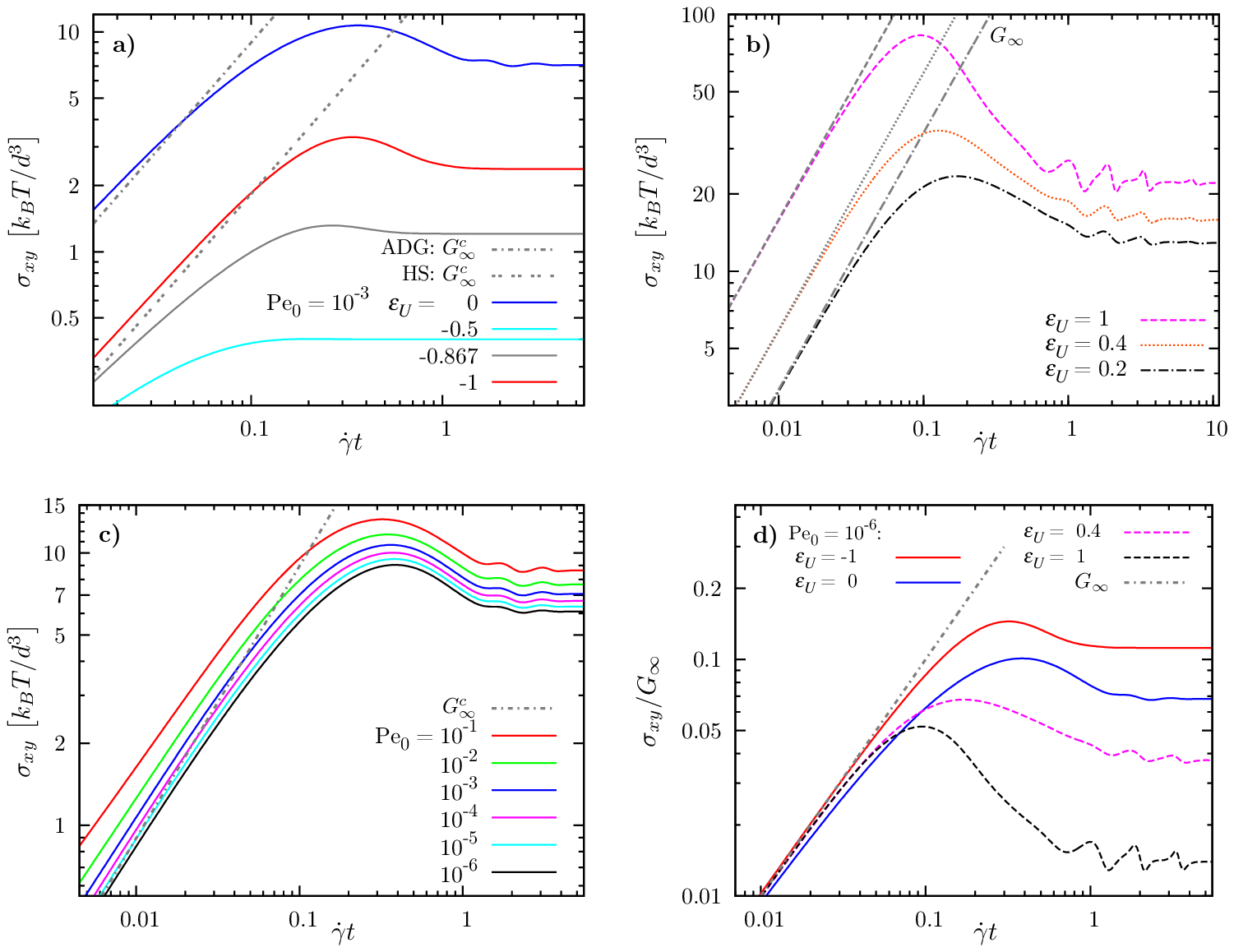}
 \caption{Overview of the transient shear stress $\sigma_{xy}$ vs accumulated strain $\dot\gamma t$ curves at fixed density  when varying the attraction strength given by the relative separation to the critical attraction strength, $\epsU=(U-U_c)/U_c$; the HS $\phi_c$ is chosen. In panel a),  $\epsU=-1, -0.867, -0.5$, and 0  correspond to the HS case, two values in the reentrant fluid region, and at the attraction-dominated transition; the shear-rate is fixed at Pe$_0=10^{-3}$. In panel b), $\epsU=0.2, 0.4$, and 1 (line style coded) enter deeper into the ADG. Here, calculations for Pe$_0$ between $10^{-6}$ and $10^{-1}$ overlap (master functions shown). 
 Note, the oscillations upon reaching the steady state at the strongest $U$ are due to the limitations of the rough $\bq$-grid discretization. Panel {\it c)} varies the shear-rate as given in the legend at the ADG-transition, $\epsU=0^+$. Panel {\it d)} shows transient stresses divided by the corresponding shear moduli, $\sigma_{xy}(t)/G_\infty$ for some glass states (labeled by attraction strength separation $\epsU$). All curves overlap in the linear regime by construction as the shear rate is low, Pe$_0=10^{-6}$.
 In all  panels, the elastic moduli $G_\infty$ and $G_\infty^c$, calculated from the plateaus of Figs.~\ref{fig::modulg3D} and \ref{figMSA::modulg3D}, or from Fig.~\ref{moduln}, are shown as {\it gray, dashed, dotted, or dash-dotted lines}.
\label{figMSA::overview}}
 \end{figure}

 Turning on shear destroys the elasticity of the amorphous solids and causes plastic deformations also in the presence of attractions. We choose the packing fraction $\phi_c$ given by the RDG  transition of hard spheres (HS)  in order to study the interplay of attractions and shear. Figure~\ref{figMSA::overview} gives an overview of the stress-strain curves for various $U$. The HS curve is included and for the solid states the elastic law with the independently calculated linear shear moduli  $G_\infty(\phi_c,U)$ from Fig.~\ref{moduln}. 
 Intermediate shear rates are chosen, in order to see strong effects but to remain well below the limit of applicability of MCT-ITT; recall that for HS, instabilities emerged for Pe$_0>10$. Starting the discussion of panel a) in Fig.~\ref{figMSA::overview} in the elastic regime and at $U=0$, the linear region in the HS stress-strain curve lies somewhat above the linear elastic asymptote $G_\infty^{\rm c, RDG} \gamma$ because the bare P\'eclet number is not asymptotically small; compare Fig.~\ref{fig::strstnFG}. 
 For attraction strengths in the reentrant fluid region, the stress-strain relations clearly exhibit fluid behavior. At $U=U_c/2$ the glassy structure on intermediate time scales relaxes so fast that the stress-strain curve becomes monotonous as holds in linear viscoelastic response, where $g_{xy}(t)$ is independent of the shear rate. We estimate Wi$=\dot\gamma\tau\approx 0.015$ there, while it is infinite for $U=0$ and $U=U_c$. For the critical attraction strength $U_c$ of the ADG, the stress-strain curve first exhibits a linear elastic regime, then an overshoot and finally a steady state. 
 The linear regime is again well described by the asymptote $G_\infty^{\rm c,ADG} \gamma$, yet, the linear shear modulus at the ADG transition exceeds the HS one by roughly a factor $4.9$. The linear elastic response holds over a far smaller strain range than for HS, and the ensuing stress peak is far broader at the ADG than at the RDG transition. The physical bonds apparently start to get broken already at strains comparable to the attraction width so that plastic rearrangements occur and the linear elastic limit is left early. The position $\gamma_*$ of the stress peak, however, is unexpectedly somewhat larger at the ADG than at the RDG transition. Yet bonds --- possibly by rotating and stretching --- still manage to bear stress up to strains even somewhat larger than characteristic for the RDG. Thus the stress overshoot is broad. Deep in the ADG, at twice the attraction strength than at the transition, the stress strain curve exhibits the same regimes. 
 Yet, the elastic modulus is larger and the position of the stress overshoot has shifted to noticeably smaller values. Also, the relative height of the overshoot has increased very strongly; see panel b) in Fig.~\ref{figMSA::overview} and Fig.~\ref{uscan} below. Consequently, the steady state stress $\sigma^\text{st}_{xy}(\gamma\to\infty)$ is higher because of the steeper linear elastic increase, but lower than this effect would imply because of the large stress release after the maximal stress $\sigma^\text{pk}_{xy}=\sigma_{xy}(\gamma^*)$. 
 This is brought out clearly in panel d) of Fig.~\ref{figMSA::overview}, where the stress divided by the shear modulus is shown for a representative set of attraction strengths. All curves overlap in the linear regime by construction. States affected by the short ranged attraction leave the linear regime at very small strains, while the hard sphere curve follows the linear elastic response far longer. The  large fraction of stress released during yielding in states deep in the ADG is apparent, and the broadness of the stress overshoots with attractions. 
 The transient stress curves at the high $U$ also exhibit a broad maximum, which approaches the long-time limit only around strains of order unity. On top of this slow transient, the calculations in Fig.~\ref{figMSA::overview} exhibit noticeable oscillations which we consider artifacts of the coarse discretization of Eq.~\eqref{eq:modmct}.

 \begin{figure}[htb]
  \centering
  \includegraphics[width=.7\linewidth]{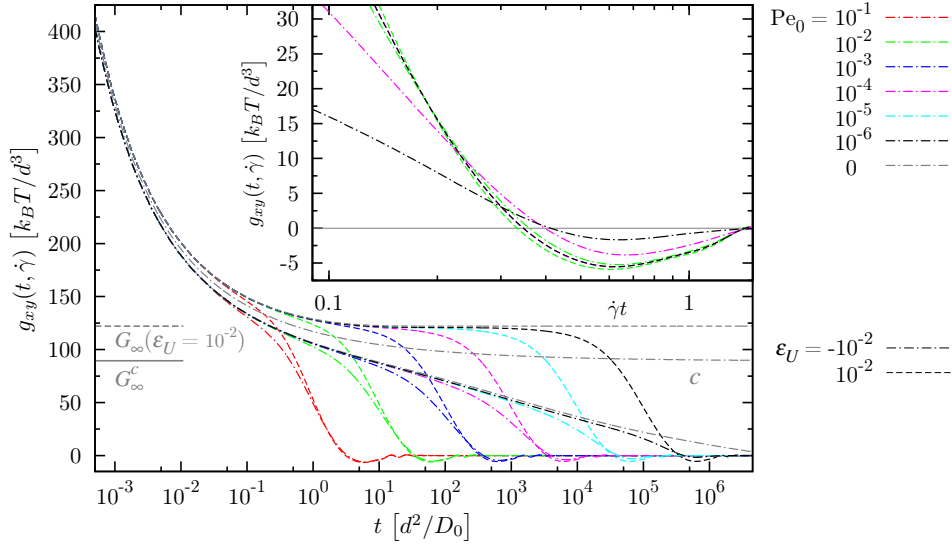}
  \caption{Generalized shear modulus $g_{xy}(t,\dot\gamma)$ of hard spheres with square-well attraction close to the ADG transition, as function of time {\it (main panel)} and accumulated strain ({\it inset}; curves for every second Pe$_0$ left out for clarity). The legend provides {\it color coded} the strain rates and {\it line-style coded} the relative attractions $\epsU=(U-U_c)/U_c$; all curves at the HS $\phi_c$. A letter $\textcolor{gray}{c}$ labels the critical $g_{xy}(t,0)$ for $\epsU=0^+$. 
  Elastic shear moduli $G_\infty(\epsU)$ can be read off from quiescent curves (Pe$_0=0$), with $G_\infty(\varepsilon=0^+)=G_\infty^c=89.5\,k_BT/d^3$ and $G_\infty(\varepsilon=10^{-2})=122\,k_BT/d^3$. The {\it inset} shows subtle differences in the peak position $\gamma_*$, where $g_{xy}(t,\dot\gamma)=0$, which are caused by $\dot\gamma$ independent $\beta$ and $\alpha$ decays; glass curves for Pe$_0=10^{-4}$ and $10^{-6}$ overlap. For Pe$_0=10^{-6}$ in the fluid,  the undershoot (almost) disappears.\label{figMSA::modulg3D}}
 \end{figure}

 After this overview of the yielding and plastic deformation when turning on a short ranged attraction, the different effects shall be explored in more detail. At first the region close to the critical attraction strength $U_c$ shall be explored varying $\epsU$ slightly, and second the behavior for large $U$ is investigated. Finally, the steady state flow curves are presented. Recall that $\epsU= (U-U_c)/U_c$, i.e. $\epsU=-1$ in the HS case, $\epsU=0$ at the ADG transition line, and  $\epsU={\cal O}(1)$ in the ADG glass region.

 Panel b) in Fig.~\ref{figMSA::overview} and Fig.~\ref{figMSA::modulg3D} present a detailed look at the transient stress evolution close to the ADG transition where shear rate, non-dimensionalized as bare P\'eclet number, and relative separation $\epsU$ are varied.
 For attraction strengths close to the ADG, the linear elastic regime generally is followed by a stress maximum. Here, the linear elasticity only holds for rather small strains. Starting at strains below  5\%, the stress grows sub-linearly with strain. As the final approach to the steady state asymptote requires strains of order unity, in general a rather broad stress overshoot can be seen in the ADG. Comparing the relative magnitude of the stress overshoot and its strain-position $\gamma_*$, similar results are observed as in the HS case.

\begin{figure}[htb]
  \centering
  \includegraphics[width=.7\linewidth]{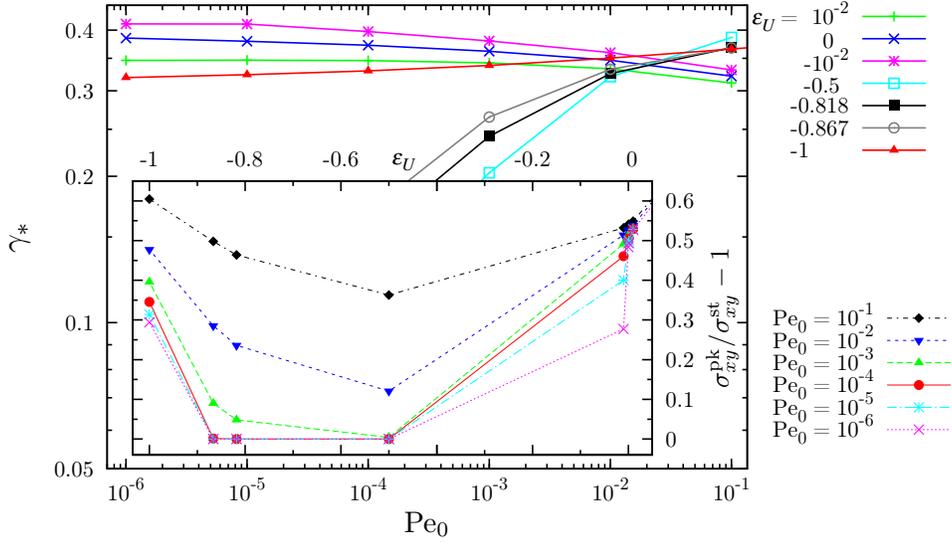}
  \caption{The {\it main panel} shows peak-strain values $\gamma_*$ as function of bare P\'eclet number Pe$_0$ going from the hard sphere glass transition ($\epsU=-1$) through the reentrant fluid state($-1<\epsU<0$)  to the  attraction dominated glass transition ($\epsU=0$); the packing fraction is fixed at $\phi_c$. The upper outside legend gives the relative separations in attraction strength $\epsU=(U-U_c)/U_c$. The $\gamma_*$s are read off from the zeros of $g_{xy}(t,\dot\gamma)$, Eq.~\eqref{eq:modmct}. 
  {\it Symbols} are connected with straight lines. 
  The {\it inset} shows the relative overshoot height $\sigma^\text{pk}_{xy}/\sigma^\text{st}_{xy}-1$ of the stress overshoots as function of the relative separation in attraction strength $\epsU$ and for several Pe$_0$ (lower outside legend). The variations mainly result from the shift of the broad $\alpha$-process through the range set by the shear rate. 
  with inverse temperature (similar to the HS case).\label{figMSA::PeakAnal3D}}
 \end{figure}

 To support this comparison, Fig.~\ref{figMSA::PeakAnal3D} shows stress-peak strains $\gamma_*$ and relative overshoot magnitudes $\sigma^\text{pk}_{xy}/\sigma^\text{st}_{xy}-1$ for different $\epsU$ and Pe$_0$ around the ADG and RDG transition. They, correspond to the zeros and negative areas in Fig.~\ref{figMSA::modulg3D}, respectively. A noteworthy and at first unexpected effect is the larger characteristic strain value at the ADG than at the RDG. For shear rates giving Pe$_0\le 0.01$, the $\gamma_*$'s at the transition are somewhat larger  for a glass where particles feel strong bonds to their neighbors than for a glass where repulsive interactions dominate. 
 Close to the ADG, the characteristic strain $\gamma_*$ does not correlate with the localization length, which is far smaller at the ADG than at the RDG as shown in the inset of Fig.~\ref{figMSA::DawsonPhases}. The cause of this subtle effect, which also depends on shear rate, is given by the extreme stretching of the quiescent $\alpha$-process at this ADG transition. The physical cause was analyzed in detail by \cite{sperl02}: The scenario of two different glass transition lines causes very broad relaxation curves, culminating in logarithmic relaxation close to higher order glass transitions. 
 At the present choice of density, attraction range and strength, the quiescent shear modulus  happens to exhibit logarithmic decay for more than five decades; see Fig.~\ref{figMSA::modulg3D}. This causes a shear-distortion of the shear modulus around its zero which pushes $\gamma_*$ to larger values. The inset of   Fig.~\ref{figMSA::modulg3D} shows the shallow negative region caused by the broad $\alpha$-process for small shear rates. Thus $\gamma_*$ increases for low Pe$_0$. 
 For larger shear rates the relaxation curves become steeper with larger $U$, and thus $\gamma_*$ decreases entering the ADG. The extremely slow intrinsic relaxation also explains the broadness of the stress-overshoot because shear affects the internal relaxation in a wide time window.  This can also be deduced from the inset of Fig.~\ref{figMSA::PeakAnal3D}, which shows that the stress-overshoot exists for a broad window of shear rates at the ADG. Again this holds because the intrinsic $\alpha$-process is characterized by a broad distribution of relaxation times. The broader crossover between a characteristically harder elastic regime and the steady state, reached after a larger plastic stress release, thus is the qualitative difference of the stress-strain curves at an attraction driven compared to a repulsion driven glass transition.

 Figure~\ref{uscan} continues the investigation of Fig.~\ref{figMSA::PeakAnal3D}, i.e.~peak strain $\gamma_*$ and relative overshoot magnitudes as function of $\epsU$ and Pe$_0$, but deep inside the ADG phase. There, the transient stress evolution changes strongly and in a characteristic way. First, the elastic constants $G_\infty$ increase dramatically. The inset of Fig.~\ref{moduln} shows that at an attraction strength twice as large as the critical value of the ADG transition, the shear constant has increased by a factor around one hundred relative to the HS one. This can also be seen directly from the linear elastic regime in the stress-strain curves in Fig.~\ref{figMSA::overview} b). Non-linearities set in at strain values comparable to the attraction range, as was observed at the ADG transition already. 
 Thus, deep inside the ADG phase where the overshoot peak is dominated completely by the elastic energy stored in the short range particle bonds (which is discussed in more detail together with Fig.~\ref{kintegral} below), a bond reordering at smaller strain values shifts the peak position to much smaller $\gamma_*$ for increasing $U$.
 The stress-overshoot retains its width in strain values, as the steady flow curve values is approached at accumulated strains of order unity (as we find for all considered states; recall that the wiggles in Fig.~\ref{figMSA::overview} at high $U$ are numerical artifacts). This implies that the steady state stress then arises from a competition of shearing and repulsion dominated caging which is not very strongly affected by attractions.
 In consequence, a very prominent feature of the stress-overshoot deep in the attraction driven glass is its relative magnitude. The lower panel of  Fig.~\ref{uscan} shows that the the relative magnitude $\sigma^\text{pk}_{xy}/\sigma^\text{st}_{xy}-1$ increases by around a factor six for the present path into the ADG. A large portion of the stress stored elastically for intermediate times is released during the late stage of the transient.  
 The flow-curve deep in the ADG is raised by roughly a decade for the present parameters relative to the HS flow curve. This increase is smaller by a decade than the attraction-driven increase of the elastic constant $G_\infty$. Because the internal $\beta$-relaxation at the states deep in the glass domain is quite rapid, no shear-rate dependence is observed in the stress-strain curves. This holds because the generalized shear modulus, which in general  depends on time and strain independently, has become a function of accumulated strain only, $g_{xy}(t,\dot\gamma)=g_{xy}(t\dot\gamma)$. Thus the stress-strain relations for different shear rates collapse onto a common master curve, which, for different $U$, are shown in Fig.~\ref{figMSA::overview} panel b). 

 \begin{figure}[htb]
  \centering
  \includegraphics[width=.7\linewidth]{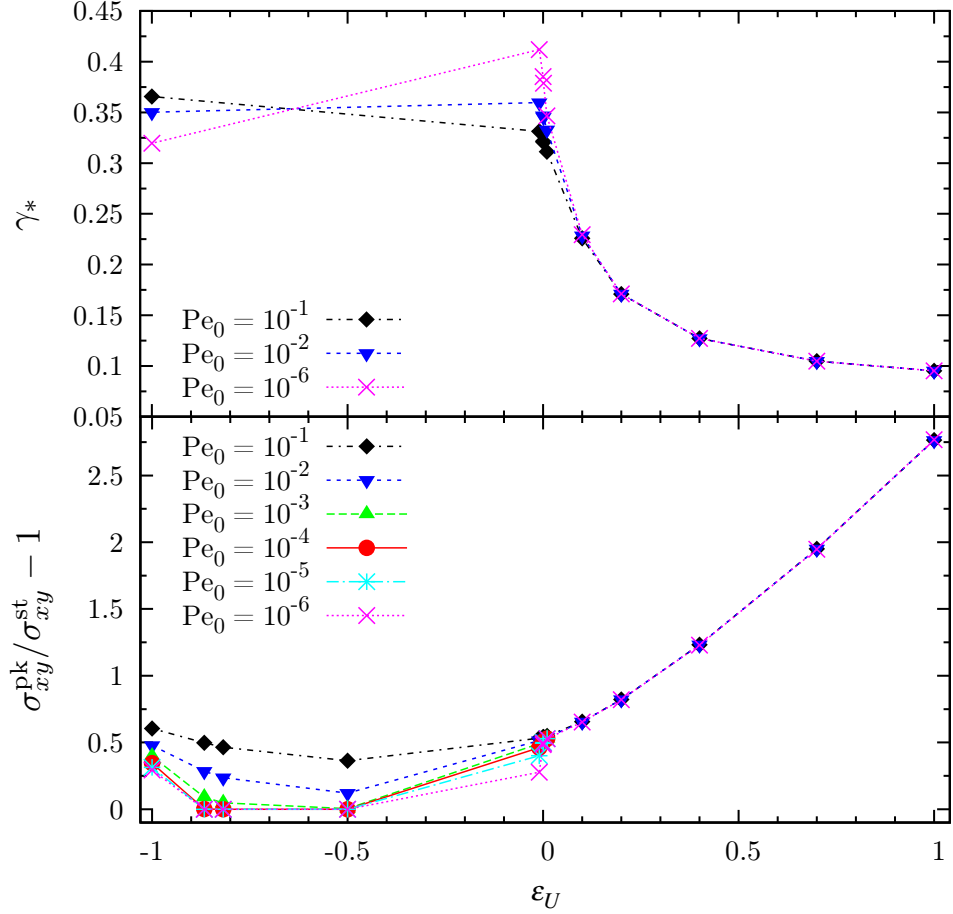}
  \caption{Peak-strain values $\gamma_*$ (upper panel)  as function of the relative separation in attraction strength $\epsU=(U-U_c)/U_c$ spanning from the hard sphere glass transition,  $\epsU=-1$, to the attraction dominated glass transition, $\epsU=0$, until deep in the ADG, $\epsU=1$. The {lower panel} shows  the relative overshoot height $\sigma^\text{pk}_{xy}/\sigma^\text{st}_{xy}-1$ of the stress overshoots as function of $\epsU$. Data for several shear rates as shown as labeled with Pe$_0$; they overlap in the explored range.\label{uscan}}
 \end{figure}

 The structural rearrangements involved in the transient stress release around $\gamma_*$ can be recognized from the wavevector dependent contributions in the generalized shear modulus. Figure~\ref{kintegral} shows the wavevector dependent vertex, viz.~the square bracket in Eq.~\eqref{eq:modmct}, for some relevant attraction strengths $U$ and densities.   The strain values are (close to) the position $\gamma_*$ of the stress overshoot. Two specific directions are chosen along the extensional ($\varphi_k=45^o$)  and compressional  ($\varphi_k=135^o$)  axis, where the distortions of the structure are maximal under shear \citep{pete12}. 
 The correlators $\Phi_{\bf q}(t)$ in Eq.~\eqref{mct1}  depend on time and accumulated strain independently. The former due to the intrinsic relaxation, the latter due to wavevector advection. In Eq.~\eqref{eq:modmct}, the squared correlators are a weight for the purely strain/advection dependent vertex, and a stress overshoot emerges depending on the different weighting of wavevector contributions. 
 If the product of structure factor deviations $S'_k S'_k(t)$ becomes negative for enough $k$ values, the $k$-space integral and thus the shear modulus become negative. At the HS transition we recover the result which \cite{zaus08} obtained by isotropic averaging that negative contributions arise close to the principal peak of the static structure factor; for HS it lies close to $kd=7$. The dominance of the principal peak in $S(k)$ at the RDG verifies that the RDG originates in the local steric packing always present in dense fluids. 
 The origin of the stress-overshoot being negative contributions from the local order peak in $S(k)$, also holds deep in the RDG, viz.~for HS at $\varepsilon=0.1$, where the vertex  has grown with density and varies more rapidly with wavevector, and where the corresponding density correlators are more glass like, viz.~have higher plateau amplitudes. Figure~\ref{kintegral} includes the vertex for  $\varepsilon=0.1$ to exemplify this; other results at this $\varepsilon$  are not shown as they can be extrapolated based on the data presented in Sect.~\ref{sec::HS}: E.g.~the characteristic strain has changed little relative to the HS transition at $\varepsilon=0$, and takes the value $\gamma_*(\varepsilon=0.1)=0.33$ at Pe$_0=10^{-6}$. 
 Somewhat unexpectedly, the same wavevector range as at the RDG transition dominates the stress integral at the stress maximum of the ADG transition. This is shown in the panel at $\epsU=0$ of  Fig.~\ref{kintegral}. One notices that larger wavevector contributions have grown, but the dominant contributions remains close to the peak in $S_q$. The large wavevector contributions in MCT-stress kernels capture the formation of physical bonds which result from the  increased stickiness of the attractive square-well potential. It increases the equilibrium structure factor at large $k$.  These high-$k$ modes are responsible for the early breakdown of the linear elastic regime in ADG states; see Fig.~\ref{figMSA::overview}. Nevertheless, the dominant contribution at the ADG transition remains connected with the local ordering of neighbor shells. 
 Apparently,  this is a universal characteristic of  the yielding process of a glass at the transition in MCT-ITT. This holds even though the elastic modulus is characteristically larger at the ADG than at the RDG as shown in Fig.~\ref{moduln}. The situation changes deep in the ADG at $U=2U_c$ (i.e. $\epsU=1$), where Fig.~\ref{kintegral} indicates that negative contributions in the stress relaxation arise dominantly at large wavevectors. Then the characteristic strain $\gamma_*$ becomes smaller, see Fig.~\ref{uscan},  and the fraction of released stress grows strongly. 
 This holds because the contributions at large wavevectors, viz.~local rearrangements of the physical bonds, are rather rapid. When bond-formation and breakage dominate the stresses deep in the ADG, the advected wavevector changes quickly with time, and the vertices of MCT, which are positive only in the quiescent state, become negative rapidly. The presence of contributions from the main peak in $S_q$ apparently cause that the bonded glass can rearrange (quasi-) elastically until the packing-dominated cages yield. Then the stress is released, which was stored elastically in the physical bonds. As the bonds become distorted starting from very small strains, a characteristically broad stress-overshoot results. The vertex at $U=2U_c$ in Fig.~\ref{kintegral} is not decayed  to zero at the upper cut-off in $k$ of our integration. 
 This indicates that the results at $\epsU=1$ are not completely converged. Yet, we expect only quantitative corrections because of the evolution of results from smaller $\epsU$. 

 \begin{figure}[htb]
  \centering
  \includegraphics[width=.9\linewidth]{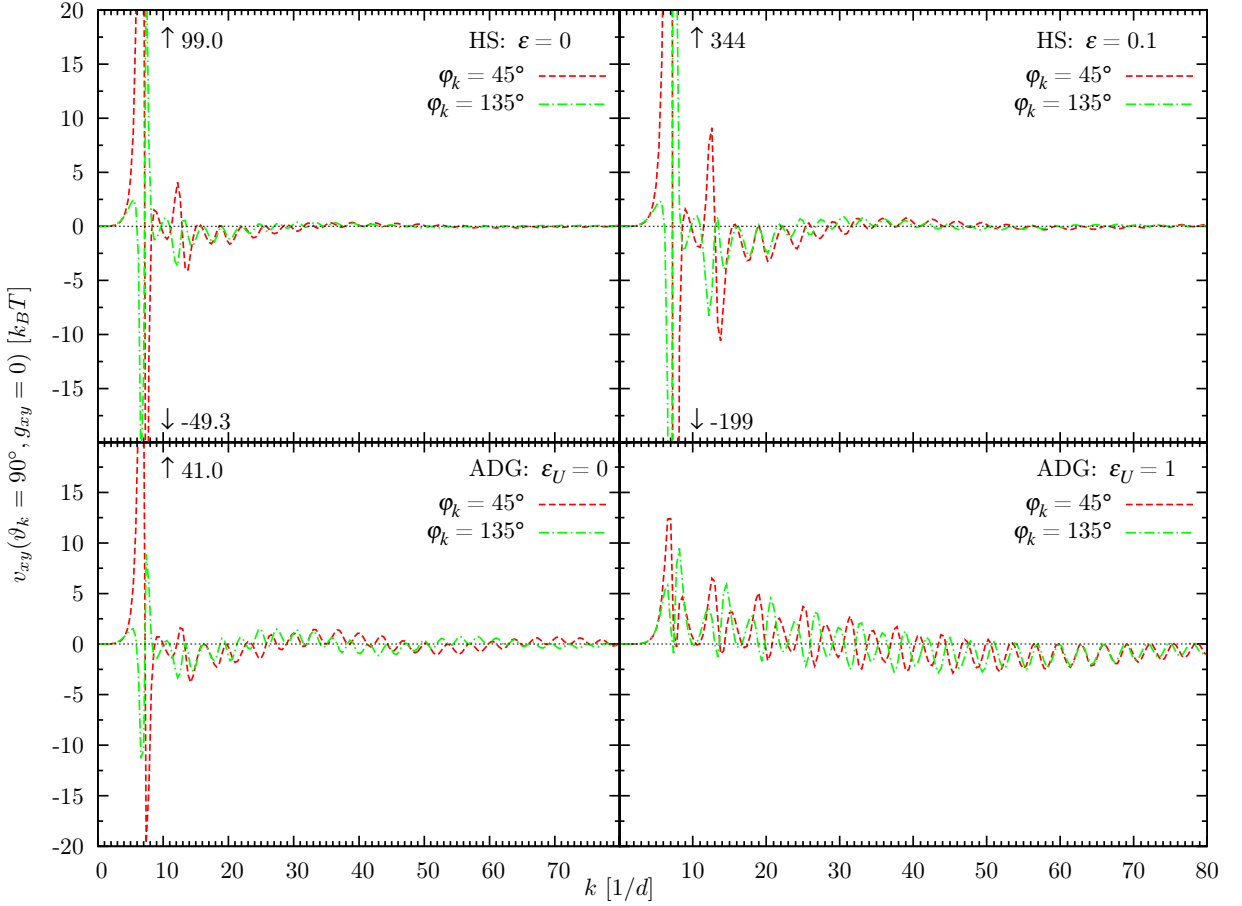}
  \caption{Wavevector $\bf k$ dependent vertex of the generalized shear modulus in Eq.~\eqref{eq:modmct}; viz.~the contents of the square bracket there.  The directions of extensional ($\varphi_k=45^o$, red lines)  and compressional  ($\varphi_k=135^o$, green lines) strain are chosen and the accumulated strain is (close) to $\gamma_*$.  Four different glasses are considered as labeled and discussed in the text. \label{kintegral}}
 \end{figure}

\begin{figure}[htb]
  \centering
  \includegraphics[width=.7\linewidth]{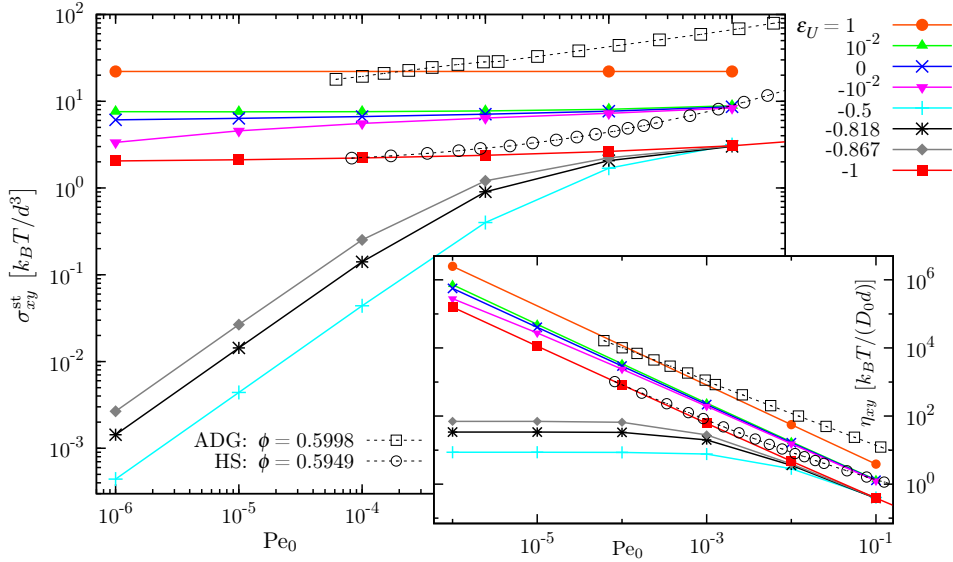}
  \caption{The {\it main panel} shows flow curves $\sigma_{xy}(t\rightarrow\infty)=\sigma^\st_{xy}$ vs bare P\'eclet number Pe$_0$ for relative attractions $\epsU=(U-U_c)/U_c$ as given in the legend. The curves span from the hard sphere glass transition,  $\epsU=-1$, to the attraction dominated glass transition, $\epsU=0$, until deep in the ADG, $\epsU=1$. The {\it colored Symbols} were calculated for six shear rates Pe$_0=10^{\{-6;\ldots;-1\}}$ and  are connected with straight lines as guides for the eye. 
  The open black symbols with  packing fractions $\phi$ given in the legend are experimental data obtained by \cite{puse08}. The comparison is discussed in  Sect.~\ref{concl}. The {\it inset} shows the stationary viscosity $\eta_{xy}=\sigma^\st_{xy}/\dot\gamma$. \label{figMSA::Stress3DFlow}}
 \end{figure}

 Figure~\ref{figMSA::Stress3DFlow} shows the flow curves for a few representative states spanning from the hard sphere to the attraction driven glass transition and beyond, deep in the ADG. The packing fraction is kept fixed at the critical value of the HS transition, so that the $\epsU=-1$ curve corresponds to the critical HS curve at $\varepsilon=0$ of Fig.~\ref{fig::Stress3DFlow}, where flow curves of hard spheres were shown. The reentrant fluid region lies at small $U$, which correspond to negative separations $\epsU<0$.  The states from $\epsU=-0.867$ to $\epsU=-0.5$ exhibit Newtonian viscosities which decrease with increasing $U$. Raising $U$ up to close to the value of the ADG transition,  the Newtonian viscosity increases again strongly. At e.g.~$\epsU=-0.01$ the Newtonian regime lies outside the window of Fig.~\ref{figMSA::Stress3DFlow} at lower bare P\'eclet numbers. 
 Thus, crossing the reentrant liquid region, the Newtonian viscosity varies non-monotonically as observed experimentally by \cite{willenbacher}. Entering the ADG, a yield stress $\sigma^+_{xy}$  arises as holds universally at MCT-ITT glass transitions. Comparing the values of $\sigma^+_{xy}$ at the HS and at the ADG transition at the same packing fraction $\phi_c$, one notices an increase caused by the  attractions. The yield stress increases by roughly the same factor as does the shear modulus $G_\infty$ at both transitions; compare Fig.~\ref{moduln}. 
 Entering into the ADG, the steady stresses increase yet again. Because the local dynamics of caging and bonding has become quite fast according to MCT-ITT deep in the ADG, the flow-curve becomes shear-rate independent. At $\epsU>0.1$ the numerical curves indicate no dependence on Pe$_0$ in the window of Fig.~\ref{figMSA::Stress3DFlow}. It is noteworthy that the increase of the steady stress in the bonded glass relative to the hard sphere glass is far smaller than the increase of the corresponding elastic constant. 
 The inset of  Fig.~\ref{moduln} indicates that $G_\infty$ has hardened by around two orders when going from $U=0$ to $U\approx10$, while the yield stress increases only from around $\sigma^+_{xy}(U=0)\approx 2k_BT/d^3$ to $\sigma^+_{xy}(U=2U_c)\approx 22.0 k_BT/d^3$.   The reason behind the comparatively weak increase in the steady stress lies in the different stress recovery after imposing flow in the ADG and the RDG. A large fraction of the stress which is elastically stored in the physical bonds  is released when the ADG fluidizes for strains of the order of $\gamma_*$; see the overview of stress-strain curves in Fig.~\ref{figMSA::overview} and the detailed analysis of the magnitude of the stress-overshoot in Fig.~\ref{figMSA::PeakAnal3D}. 

 Converting the flow curves into viscosities, $\eta_{xy}=\sigma_{xy}/\dot\gamma$, straightens out the curves over a wide range along the ordinate. The subtle sigmoidal shape of the flow curves of MCT-ITT close to a glass transition thus get ironed-out. The inset of  Fig.~\ref{figMSA::Stress3DFlow} shows the corresponding viscosities which exhibit a Newtonian plateau in fluid states for small Weissenberg numbers and then cross over to shear-thinning with asymptotic exponent -1. Restrictions in the numerical code prevent us also for the ADG to address the question of the existence of a second Newtonian plateau at high shear rates. The MCT-ITT calculations continue to decrease for the numerically accessible range of $\dot\gamma$.

\begin{figure}[htb]
  \includegraphics[width=\linewidth]{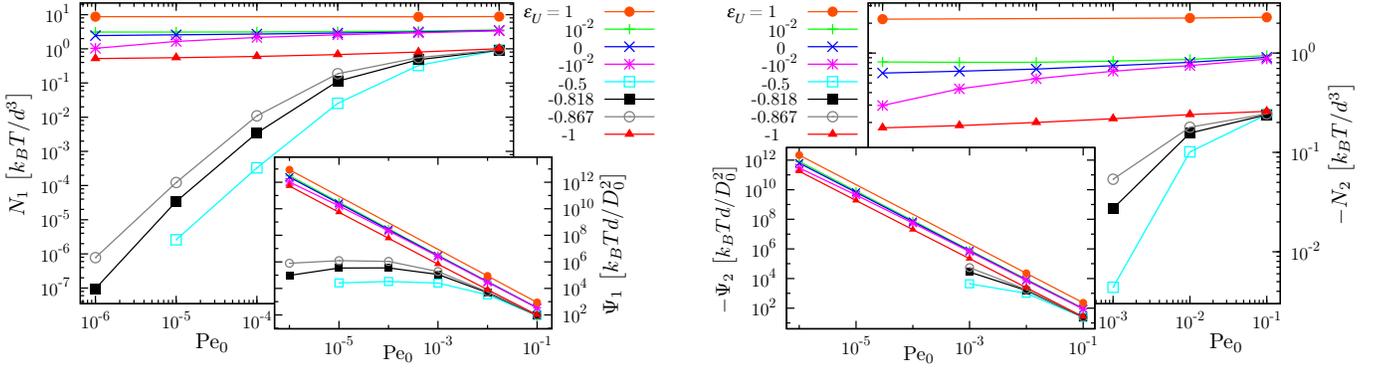}
  \caption{The {\it left main panel} shows stationary first normal-stress differences $N_1(t\rightarrow\infty)=\sigma^\st_{xx}-\sigma^\st_{yy}$ vs bare P\'eclet number Pe$_0$ for six shear rates Pe$_0=10^{\{-6;\ldots;-1\}}$ and for relative attractions $\epsU=(U-U_c)/U_c$ as given in the legend. The {\it Symbols} are connected with straight lines as guides to the eye. The {\it left inset} shows the first long-time normal-stress coefficient $\Psi_1=(\sigma^\st_{xx}-\sigma^\st_{yy})/\text{Pe}^2_0$. 
  The {\it right main panel} shows stationary second normal-stress differences $-N_2(t\rightarrow\infty)=\sigma^\st_{zz}-\sigma^\st_{yy}$, and its {\it inset} the corresponding long-time normal-stress coefficient $-\Psi_2=(\sigma^\st_{zz}-\sigma^\st_{yy})/\text{Pe}^2_0$.
\label{figMSA::N13DFlow}}
 \end{figure}

 Figure~\ref{figMSA::N13DFlow} shows the first and second normal-stress differences  $N_1=\sigma_{xx}-\sigma_{yy}$, $N_2=\sigma_{yy}-\sigma_{zz}$. Choices in the numerical algorithm aimed at capturing strong flows give errors in the coefficients when they become too small in fluid states. Hence we cannot compare with the calculations by \cite{brad13} who considered the prefactor of the quadratic scaling at low P\'eclet numbers. As in the case of the shear stress, the steady state normal stresses deep in the ADG are above the ones of the HS at the same packing fraction. 
 Yet, during the transient, again a large amount of the stress built-up during the linear elastic response is released upon yielding. The build-up of normal stresses during the deformation of physical bonds in the linear regime  again obeys the Lodge-Meissner relationship as tested in Fig.~\ref{fig::Stress3DN1N2} for hard spheres; the tests for the ADG states are not shown.

\section{Conclusions and comparison with experimental data}\label{concl}

 We presented the first fully quantitative solutions of the MCT-ITT equations for the nonlinear response of shear driven colloidal dispersions.  The results on the transient shear stress response, the steady flow curves, and normal stresses at colloidal glass transitions exhibit the central qualitative features which have previously been discussed using schematic MCT-ITT models. They recover and quantify phenomena like the initial elastic response, which is linear in the accumulated strain,  the yielding of glasses characterized by a dynamic yield stress, and the transient stress overshoot, which defines a strain $\gamma_*$ characterizing the yielding process. Solutions of the MCT-ITT equations by \cite{fuch09}, \cite{krue11}, and \cite{aman13} considering hard disks in two dimensions showed that these phenomena are universal at MCT-ITT glass transitions under shear and also arise in  two-dimensional systems confined to a plane. 
 These authors also discussed transient density correlators, tagged particle motion, and distorted structures, which, for three dimensions, can only  be presented in future. The present three-dimensional solutions, however,  enable comparison with experimental data, which are available for dispersions of colloidal hard spheres and of colloids mixed with non-adsorbing polymers which induce a short-ranged attraction of Asakura-Oosawa form.

 A quite stringent comparison of the theoretical results for hard spheres in Sect.~\ref{sec::HS} can be made with the experiments by \cite{cras08} who studied core-shell particles consisting of a polystyrene core and a thermosensitive,  crosslinked PNIPAM shell. The microgel dispersions with size polydispersity of 9.3\% can be mapped onto the phase diagram of monodisperse hard spheres using the freezing density $\phi_F=0.494$, and exhibit the strongest tendency to crystallize around $\phi_{\rm eff}=0.55$. The rheology at higher packing fractions closer to vitrification appears little affected by crystallization. The effective packing fraction can be adjusted by changing the weight percentage of particles or by changing the effective size $R_H$ (viz.~the hydrodynamic radius taken to be $d/2$ here) by temperature, which makes it possible to approach the glass transition packing fraction of hard spheres $\phi_c=0.58$ quite closely. 
 Moreover, the frequency-dependent linear response moduli $G'(\omega)$ and $G''(\omega)$ were already quantitatively analyzed using MCT so that the mapping of the theory onto the  rheology is known. \cite{cras08}  found that roughly the same value of the critical packing fraction as established by  \citep{puse87} studying PMMA hard sphere colloids using dynamic light scattering  rationalizes the microgel rheology. The separation parameters obtained by \cite{cras08} and listed in the caption of Fig.~\ref{fig::Stress3DFlow} refer to this experimental hard sphere glass transition density. Clearly, the theoretical results, which are calculated (not fitted) for comparable separation parameters from the theoretical $\phi_c$ show quite comparable flow curves.  The need to use the separation from the critical packing fraction instead of the actual value of the packing fraction in order to compare MCT and experiment is well established \citep{goetze}. 
 It arises from the approximate nature of MCT which misses the precise value of $\phi_c$, while capturing the sensitive dependence of the viscoelasticity on the separation to the glass transition. The comparison in  Fig.~\ref{fig::Stress3DFlow} can in principle be done  without adjustable parameter, because $D_0$ sets the time scale  in Eq.~\eqref{mct1} and can be calculated from the solvent viscosity following Stokes, Einstein and Sutherland \citep{eins05,suth05}. Also the stresses are calculated directly. 
 Yet, hydrodynamic interactions are neglected by MCT-ITT. They affect the short time diffusion coefficient, which at high concentrations is a more relevant  scale than the Stokes-Einstein-Sutherland diffusion coefficient at infinite dilution. Additionally, hydrodynamic interactions add a contribution to the viscosity at high shear rates.  \cite{cras08} found that the linear moduli agree best when assuming that the hydrodynamic interactions slow down the short time diffusion to $0.15 D_0$. 
 (The  viscosity $\eta_\infty^{\dot\gamma}$ observed at high shear rates is indicated in Fig.~\ref{fig::Stress3DFlow} to describe the second origin of hydrodynamic deviations.) 
 Also they observed that MCT underestimates stresses by 40\% (a rescaling factor $c_y^{G}=1.4$ was used). The data in Fig.~\ref{fig::Stress3DFlow} were rescaled by the given ratio of the diffusion coefficients, but by a different stress-rescaling  factor: $c_y^{\sigma}=0.55$.  Quite satisfactorily MCT deviates by less than 50\% from either experiment. The aspect that theory underestimates the linear elastic stress but overestimates the steady state stress of the yielding glass can be traced back to the error of MCT-ITT in determining the characteristic strain value $\gamma_*$. While transient stress-strain curves are not available for the microgel dispersions by \cite{cras08}, stress-overshoots were measured in more polydisperse microgel samples. 
 \cite{aman13} measured $\gamma_*^{\rm ex}\approx0.10$ while our MCT-ITT calculation gives $\gamma_*^{\rm mct}\approx 0.32$.  Apparently, MCT-ITT underestimates the speed-up by shear of stress fluctuations. 
 While experiments close to the glass transition in hard sphere colloids, including the large amplitude oscillatory measurements by \cite{puse02}, find characteristic strain values around 10\%, MCT-ITT overestimates it by a factor around three. Consequently the steady stresses are somewhat overestimated, even though the linear elastic regime is somewhat underestimated. 
 Reassuringly, the deviations by MCT-ITT in three dimensions are appreciably smaller than the deviations in two dimensional hard disk systems, where \cite{fuch09} found larger correction factors of the flow curves; they found  around $c_y^{\sigma}=0.1$. Considering the density dependence of the stress-overshoot peak-strain $\gamma_*$, one notices a similar increase than found by  \cite{puse02} for the strain where irreversible rearrangements first appear in states close to the glass transition. 
 Intriguingly, their light scattering  measurements  of this characteristic strain reveal a maximum at intermediate densities followed by a decrease when approaching random close packing. 
 The dependence of $\gamma_*$ on packing fraction thus is richer than the dependence of the localization length on $\phi$. The latter is expected to be monotonically decreasing with higher packing fraction as particles get localized more tightly. While both length scales thus characterize the cage effect in repulsion dominated glass transitions, and take comparable values right at the hard sphere glass transition, their precise relation is non-linear and not straightforward. 

 Turning on the short-ranged attractions by adding polymer to the colloids, first the states diagram can be tested. It consists of two different glass states, a reentrant fluid region, and possibly glass-to-glass transitions and higher order glass singularities. \cite{pham02} and \cite{bart02}  found qualitative agreement concerning the transition lines, with the ones from theory shifted, but tracking the experimental ones. \cite{willenbacher} extended these studies by pushing the reentrant fluid region  to higher packing fractions. The enhanced elastic stiffness of the glasses with physical bonds was convincingly seen by \cite{pham06} and \cite{puse08}.  While stress vs strain curves after shear start-up are not available, \cite{puse08} present and discuss as equivalent  stresses after step strain deformations. The observed characteristic strain $\gamma_*\approx 0.1$ for hard spheres corresponds well to the above discussion. 
 For glasses with a polymer-induced attraction of roughly 6\% range, the stress-strain relations show two characteristic differences to the ones of hard spheres. First, nonlinear deviations to the linear elastic response set in at rather small strain values. This, considering the differences in attraction potential, agrees well with our findings from MCT-ITT. The origin of the nonlinearities lies in the high-wavevector contributions of the memory kernels which cause a broadly stretched quiescent structural relaxation. They also are very susceptible to wavevector advection, the mechanism by which shear affects the structural relaxation in MCT-ITT. Thus small strains suffice to soften the elastic response. The second experimental finding is a second maximum of the stress at strain values beyond one, which is not observed by MCT-ITT. 
 The stress maximum at large strains is higher than the maximum at strains comparable to the repulsive case, distorting it to a shoulder. While MCT-ITT appears to capture the phenomena at the first characteristic strain  $\gamma_*$ which remains close to the hard sphere value, the second stress maximum is missed.  
 Presumably it arises from structural correlations in the bonded glassy state which reach beyond the cage-effect length scale. Going to the final steady state, Fig.~\ref{figMSA::Stress3DFlow}  contains the experimental flow curves obtained by \cite{puse08} for a repulsion and an attraction dominated glass state. The experimental data are mapped onto the theoretical calculations by estimating the diffusion coefficient  $D_0 = k_B T / (3\pi \eta_{\rm solv} d)$ by using $\eta_{\rm solv} = 1$mPa s with the particle size  $d = $260nm.  For the PMMA hard sphere colloids in an organic solvent a slightly different stress-rescaling factor needs to be used than in Fig.~\ref{fig::Stress3DFlow}. 
 The value $c_y^\sigma = 3.05$ gives the best agreement for the hard sphere data. 
 After this mapping of the hard sphere data,  the increase of the yield stress deep in an attraction dominated glass state can be addressed in Fig.~\ref{figMSA::Stress3DFlow}. While the differences in packing fraction, attraction range and strength prevent a more detailed comparison, the hardening of the flow curves at (roughly) fixed packing fraction upon increasing the attraction strength agrees qualitatively with the MCT-ITT calculation. A more detailed comparison appears justified, which would require improved equilibrium structural input. 

 Based on the encouraging comparisons of the nonlinear stress strain relations from microscopic MCT-ITT with experiments on model colloidal systems it appears worthwhile to consider shear distorted structure and transient density correlations in order to gain a deeper understanding of the mechanisms of plastic deformation and yielding of colloidal glasses. Work along these lines is underway.

\section{Acknowledgements}
We thank Th. Voigtmann for valuable discussions, M. Siebenb\"urger for help in interpreting the  experimental data, and the Deutsche Forschungsgemeinschaft (DFG) for financial support in the initiative FOR 1394, project P3.

\appendix

\section{Numerical implementation\label{sec::implementation}}
 
 In this appendix, the numerical implementation of MCT-ITT in three dimensions is summarized. The standard challenge to solve MCT-equations over around ten decades in time  is made more difficult by the requirement to compute $2\cdot d$-dimensional wavevector integrals for the memory kernels. Desirable requirements to the numerics are to recover an isotropic quiescent solution \citep{singh97}, while choosing a sufficiently close $\bq$-grid discretization in Fourier space, in order to minimize discretization errors. 
 
 The numerical evaluation of Eq.~\eqref{mct1} depends mainly on the discretization of the friction kernel $m_{\bf q}(t,t')$ in Eq.~\eqref{mct2}. Under simple start-up shear, the dependence on two times $t$ and $t'<t$  can be simplified on the dependence on the time interval $\tau=t-t'$, yielding $m_{\bf q(t')}(\tau)$, which is explained in detail by \cite{cate09}. This yields a so called single-time MCT, which is far simpler to solve than two-times MCT considered by \cite{voigt12}. The temporal evaluation of Eq.~\eqref{mct1} follows the standard scheme of previous MCT-ITT calculations (also schematic MCT) and is described in detail e.g.~by \cite{voigt12} and \cite{Amann2013}. 
 The temporal discretization is performed on a time grid consisting of blocks, which each consist of $N_t=64$ linearly spaced time instances. A straightforward discretization backwards in time  (see Brader-Voigtmann algorithm \citep{Amann2013}) and $\Phi_{\bf q}(t)$ is iteratively solved, depending on all time instances $t'\leqslant t$. 
 The first time block is initialized by ${\Phi}_{\bf q}=\exp(-\Gamma_{\bf q} t)$. As second input, the Percus Yevick (PY) structure factor for hard spheres and square well potentials is used \citep{hansen,baxt70,daws00}. Subsequent time blocks are generated by doubling the  time step  and using the arithmetic average of $\Phi_{\bf q}$ and $m_{\bf q}(t)$ of two instances of the preceding time block to initialize the first $N_t/2$ instances of the new time block. Via this decimation procedure, a large time range can be covered. 

 The discretization of wavevectors in $\bq$ space (and the $\bk$ integral in the friction kernel, Eq.~\eqref{mct2}) has been performed using standard spherical coordinates, $\bs{q}=\left(q_x\\q_y\\q_z\right)=q\left(\cos\varphi_q\sin\vartheta_q \sin\varphi_q\sin\vartheta_q\\ \cos\vartheta_q\right)$, discretizing modulus $q$, azimuthal angle $\varphi_q$, and inclination angle $\vartheta_q$. This helps to keep the isotropy in quiescent calculations and simplifies identifying spherical symmetries in the computed observables. Under shear however, the $\bq$ grid becomes completely anisotropic in three dimensions.  The $V_\mathbf{qkp}(t)$ is $2\cdot d$ dimensional. The computation of $\bk(t)$ respects shear advection via coordinate transformations and depends on the according $\bq(t=0)$, which has been chosen as $k_z$ axis; see \cite{Amann2013} for more details.
 Table~\ref{tab::main_grid} shows the generic parameter choice used for the computation of the results of this work as best trade-off between computation time and precision.
 \begin{table}[htb]
 \centering
  \begin{tabular}[c]{c|c|c|c|c||c}
         HS $q$ & ADG $q$ & $\Delta q$ & $2\pi/\Delta \varphi_q$ & $\pi/\Delta \vartheta_q$ & HS $\phi_c$\\
  \hline [0.2;39.8] & [0.2;79.8] & 0.4 & 24 & 24 & 0.515712(1)
  \end{tabular}\\[2mm]
 \caption[Preferred $\bq$ grid parameters in Chap.~\ref{chap::MCT3D}]{Generic parameters of the $\bq$ grid used for this work. $\phi_c$ is rounded up in the seventh digit.\label{tab::main_grid}}
 \end{table}

 With {\it OpenMP} parallelization on 32 CPUs \'a $2.6\,$GHz and $66\,$GB RAM, the calculation of one stress-strain curve in the ADG takes up to 90 hours. Only the repeated calculation of $m_{\bf q(t')}(\tau)$ can be parallelized effectively to gain computation speed. Hence, a compromise in precision and computation time must be accepted.  A method of saving computation time in the iteration of $\Phi_{\bf q}(t)$, which needs a repeated calculation of $m_{\bf q(t')}(\tau)$, is to store the vertex for the youngest time instance $t$, which consumes a relevant fraction of the available RAM and limits the grid discretization. Thus, RAM consumption and computation time increase approximately with the square of the grid-point density.

\clearpage


\end{document}